\documentstyle{article}

\begin{document}

\newtheorem{theorem}{Theorem}
\newtheorem{definition}{Definition}
\newtheorem{deflem}{Definition and Lemma}
\newtheorem{lemma}{Lemma}
\newtheorem{example}{Example}
\newtheorem{remark}{Remark}
\newtheorem{remarks}{Remarks}
\newtheorem{cor}{Corollary}

\def\N{{\rm I\kern-.1567em N}}                              
\def\No{\N_0}                                               
\def\R{{\rm I\kern-.1567em R}}                              
\def\C{{\rm C\kern-4.7pt                                    
\vrule height 7.7pt width 0.4pt depth -0.5pt \phantom {.}}}
\def\Q{{\rm Q\kern-5.7pt                                    
\vrule height 6.5pt width 0.4pt depth -0.5pt \phantom {.}}}
\def\Z{{\sf Z\kern-4.5pt Z}}                                

\newcommand{\eps}{\varepsilon}
\newcommand{\dt}{\partial_t}
\newcommand{\dx}{\nabla_x}
\newcommand{\dv}{\nabla_v}

\title{The Vlasov-Poisson system with radiation damping}
\author{{\sc Markus Kunze$^{1}$\& Alan D. Rendall$^{2}$}\\[2ex]
        $^{1}$ Zentrum Mathematik, TU M\"unchen, \\
        Gabelsbergerstr.~49, D\,-\,80333 M\"unchen, Germany\\
        e-mail: mkunze@mathematik.tu-muenchen.de \\[1ex]
        $^{2}$ Max-Planck-Institut f\"ur Gravitationsphysik,\\
        Am M\"uhlenberg 1,
        D\,-\,14476 Golm, Germany \\
        e-mail: rendall@aei-potsdam.mpg.de \\[2ex]
        {\bf Key words:} radiation damping, Vlasov-Poisson system, 
        \\ global existence, long-time behaviour, runaway solutions}
\date{}
\maketitle
\begin{abstract}
\noindent
We set up and analyze a model of radiation damping within the framework
of continuum mechanics, inspired by a model of post-Newtonian
hydrodynamics due to Blanchet, Damour and Sch\"{a}fer. In order to simplify
the problem as much as possible we replace the gravitational field by
the electromagnetic field and the fluid by kinetic theory. We prove
that the resulting system has a well-posed Cauchy problem globally in time
for general initial data and that in all solutions the fields decay to zero 
at late times.
In particular, this means that the model is free 
from the runaway solutions which frequently
occur in descriptions 
of radiation reaction.
\end{abstract}


\hspace{1em}\\[-8ex] 

\section{Introduction and main results}

The Vlasov-Poisson system is a well-known description of collisionless
particles which interact via a field which they generate collectively.
It can be applied in the case of particles interacting through the
electromagnetic field (plasma physics case) or the gravitational field
(stellar dynamics case). The equations modelling the two cases are
only distinguished by a difference of sign. This description is
non-relativistic and is only appropriate for physical situations where
the velocities of the particles are small compared to the velocity of
light. When it is replaced by a fully relativistic model the two
cases diverge drastically. In the electromagnetic case the appropriate
system of equations is the (relativistic) Vlasov-Maxwell system while in
the gravitational case it is the Vlasov-Einstein system, which is much
more complicated.

In classical electrodynamics it is well known that accelerated charged
particles radiate and that this leads to an effect on the motion of the
particles known as radiation reaction. This typically leads to damping,
i.e.~to loss of energy by the particles. A similar but more complicated
effect occurs in the case of the gravitational field. It is, however,
hard to formulate exactly due to difficulties such as the nonlinearity
and coordinate dependence of the equations used. There is a large
literature concerning effective equations in electrodynamics which
incorporate radiation damping without providing a full relativistic
description of the field and sources. These effective equations usually
have undesirable solutions which tend to infinity exponentially fast,
the so-called ``runaway solutions''. It has recently been observed
that nevertheless, in some of these models, the physically relevant
solutions of the effective equation constitute a center-like manifold
in phase space, restricted to which the dynamics is completely well-behaved.
Moreover, the effective equation is a good approximation of the
full system; cf.~\cite{KS1,KS2}.

In the case of the gravitational field, radiation damping is a subject of
particular interest at the moment due to the fact that gravitational
wave detectors will soon be ready to go into operation and it is
important for their effective functioning that the sources of gravitational
waves be understood well. (For background on gravitational wave detection
see for instance \cite{flanagan} and references therein.) The most
promising type of source at the moment is a strongly self-gravitating
system of two stars rotating about their common centre of mass which
lose energy by (gravitational) radiation damping and eventually coalesce.
As has already been indicated, it is hard to describe this within the
full theory and hence effective equations like those known in electrodynamics
are important. Very little is understood about this in terms of rigorous
mathematics at this time. The aim of this paper is to take a first step
towards bringing this subject into the domain where models can be defined
in a mathematically precise way and theorems proved about them.

The model we will discuss has the following characteristics. It clearly
exhibits the phenomenon of radiation damping. It does not suffer from
pathologies such as runaway solutions. It is simple enough so that we
can prove theorems about the global behaviour of the general solution.
The particular model was chosen with the aim of obtaining this combination
of properties. It is inspired by a model of Blanchet, Damour and Sch\"afer
\cite{bds} for a perfect fluid with radiation damping. The phenomenon of
radiation damping is intimately connected with the long time asymptotics of
the system. Thus, in order to capture it mathematically, we need at least
a global existence theorem. This seems hopeless for a fluid, due to the
formation of shocks, and so we replace it by collisionless matter
(Euler replaced by Vlasov). The latter is known to have good global existence
properties \cite{pfaff,lio-pert,schaeff,horst}. Although the original motivation
came from the gravitational case, the electromagnetic case is much simpler.
Thus we use a model motivated by the electromagnetic case here, hoping to
return to the more complicated gravitational case at a later date. We are
not aware that the model used here has a direct physical application.

\hspace{1em}\\[3ex]
The model to be studied is defined as follows. There are two species of
particles of opposite charges, say ions (``$+$'') and electrons (``$-$'').
In the case of the Vlasov-Poisson system the motion of the individual
particles is governed by the characteristic systems
\begin{eqnarray}
   \dot{X}^+ & = & V^+,\quad \dot{V}^+=\nabla U(t, X^+),
   \label{cl-char1} \\[1ex]
   \dot{X}^- & = & V^-,\quad \dot{V}^-=-\nabla U(t, X^-),
   \label{cl-char2}
\end{eqnarray}
where $U=U(t, x)$ is the (electric) potential. The requirement that
the particle densities $f^{\pm}=f^{\pm}(t, x, v)$ be constant
along the characteristics leads to the Vlasov equations
\begin{eqnarray}
   \dt f^+ + v\cdot\dx f^+
   +\nabla U\cdot\dv f^+ & = & 0, \label{class-vlas1} \\[1ex]
   \dt f^- + v\cdot\dx f^-
   -\nabla U\cdot\dv f^- & = & 0, \label{class-vlas2}
\end{eqnarray}
with $t\in\R$, $x\in\R^3$, and $v\in\R^3$ denoting time,
position, and velocity variable, respectively.
The potential $U$ derives from the Poisson equation
\begin{equation}\label{DelU}
   \Delta U=4\pi\rho=4\pi (\rho^+ -\rho^-),
   \quad \lim_{|x|\to\infty}U(t, x)=0,
\end{equation}
where
\begin{equation}\label{rho-def}
   \rho^{\pm}(t, x)=\int f^{\pm}(t, x, v)\,dv.
\end{equation}
Supplied with suitable data $f^{\pm}(t=0)=f_0^{\pm}$,
(\ref{class-vlas1})--(\ref{rho-def}) constitutes the Vlasov-Poisson system
for two species of opposite charges; see \cite{glassey,gerh}
for general information on Vlasov-Poisson and related models.

In order to introduce a damping effect due to radiation into
(\ref{class-vlas1})--(\ref{rho-def}), we modify the characteristic
equations by introducing a small additional term. Let
\begin{equation}\label{D-def}
   D(t)=\int x\rho(t, x)\,dx=\int\int x(f^+(t, x, v)
   -f^-(t, x, v))\,dx dv
\end{equation}
denote the corresponding dipole moment, and replace
(\ref{cl-char1}), (\ref{cl-char2}) by
\begin{eqnarray}
   \dot{X}^+ & = & V^+,\quad \dot{V}^+=\nabla U(t, X^+)
   +\eps\stackrel{...}{D}(t), \label{char2} \\[1ex]
   \dot{X}^- & = & V^-,\quad \dot{V}^-=-\nabla U(t, X^-)
   -\eps\stackrel{...}{D}(t), \label{char1}
\end{eqnarray}
with an $\eps>0$ small. This is to be thought of as an approximation to
the full Vlasov-Maxwell system. It includes the electric dipole radiation
which is supposed to give the leading contribution to the radiation
reaction, cf.~\cite[p.~784]{jackson}.  

The third time derivative in these equations can lead to pathological
behaviour and so we will modify the model by formally small corrections
so as to eliminate it. Here we follow the procedure of \cite{bds} which
was used to tackle the fifth time derivatives which occur in the
analogous gravitational problem. To reduce the order of derivatives on
$D(t)$, we utilize
the transformations
\begin{equation}\label{abl-tr}
   \tilde{V}^+=V^+ -\eps\ddot D(t)
   \quad\mbox{and}\quad
   \tilde{V}^-=V^- +\eps\ddot D(t).
\end{equation}
Then (\ref{char2}), (\ref{char1}) read
\begin{eqnarray}
   \dot{X}^+ & = & V^+ +\eps\ddot{D}(t),\quad
   \dot{V}^+=\nabla U(t, X^+), \label{char4} \\[1ex]
   \dot{X}^- & = & V^- -\eps\ddot{D}(t),
   \quad \dot{V}^-=-\nabla U(t, X^-), \label{char3}
\end{eqnarray}
where the tilde has been omitted for simplicity.
The corresponding Vlasov equations are then
\begin{eqnarray}
   \dt f^+ + (v+\eps\ddot{D}(t))\cdot\dx f^+
   +\nabla U\cdot\dv f^+ & = & 0, \label{vlas2} \\[1ex]
   \dt f^- + (v-\eps\ddot{D}(t))\cdot\dx f^-
   -\nabla U\cdot\dv f^- & = & 0. \label{vlas1}
\end{eqnarray}
Next we derive an approximation $D^{[2]}(t)$ to $\ddot{D}(t)$.
By definition of $D(t)$ in (\ref{D-def}) we formally calculate
\begin{eqnarray*}
   \dot{D}(t) & = & \int\int x (\dt f^+ -\dt f^-)\,dx dv
   \\ & = & {\cal O}(\eps)+\int\int x
   \Big(-v\cdot\dx f^+ -\nabla U\cdot\dv f^+
   +v\cdot\dx f^- -\nabla U\cdot\dv f^-\Big)\,dx dv
   \\ & = & {\cal O}(\eps)+\int\int v (f^+-f^-)\,dx dv.
\end{eqnarray*}
Thus
\begin{eqnarray*}
   \ddot{D}(t) & = & {\cal O}(\eps)+\int\int
   v (\dt f^+ -\dt f^-)\,dx dv
   \\ & = & {\cal O}(\eps)+\int\int v\Big(-v\cdot\dx f^+
   -\nabla U\cdot\dv f^+
   +v\cdot\dx f^- -\nabla U\cdot\dv f^-\Big)\,dx dv
   \\ & = & {\cal O}(\eps)-\int\int v\,\nabla U\cdot
   (\dv f^+ +\dv f^-)\,dx dv
   ={\cal O}(\eps)+\int\int\nabla U (f^+ + f^-)\,dx dv.
\end{eqnarray*}
Hence we are led to define the approximation
\begin{eqnarray}\label{d2def}
   D^{[2]}(t) & = & \int\int \nabla U(t, x) (f^+(t, x, v)
   + f^-(t, x, v))\,dx dv \nonumber \\ & = &
   \int\nabla U(t, x) (\rho^+(t, x)+ \rho^-(t, x))\,dx,
\end{eqnarray}
and we replace (\ref{char4}), (\ref{char3}) by
\begin{eqnarray}
   \dot{X}^+ & = & V^+ +\eps D^{[2]}(t),\quad
   \dot{V}^+=\nabla U(t, X^+), \label{char6} \\[1ex]
   \dot{X}^- & = & V^- -\eps D^{[2]}(t),
   \quad \dot{V}^-=-\nabla U(t, X^-), \label{char5}
\end{eqnarray}
with corresponding Vlasov equations
\begin{eqnarray}
   \dt f^+ + (v+\eps D^{[2]}(t))\cdot\dx f^+
   +\nabla U\cdot\dv f^+ & = & 0, \label{vlas+} \\[1ex]
   \dt f^- + (v-\eps D^{[2]}(t))\cdot\dx f^-
   -\nabla U\cdot\dv f^- & = & 0, \label{vlas-}
\end{eqnarray}
and $U$ is determined by (\ref{DelU}).

We call the system consisting of (\ref{vlas+}), (\ref{vlas-}),
(\ref{DelU}), (\ref{rho-def}), and (\ref{d2def}) the
Vlasov-Poisson system with damping (VPD), and we propose it
as a model to study the damping effect due to radiation.
We add some more comments.

\begin{remark}{\rm (a) To model radiation as we have done it,
it is necessary to consider at least two species with different
charge to mass ratios. Here we make the simplest choice of
equal masses and two charges which are equal in magnitude and opposite
in sign.
If the charge to mass ratios were equal then the rate of change of the
dipole moment would be proportional to the linear momentum of the system
and, by conservation of momentum, the radiation reaction force would
vanish. This is a well-known fact (absence of bremsstrahlung for
identical particles), cf.~\cite[p.~411]{burke} or \cite[p.~201]{zhelez},
and can also be seen from the corresponding effective equations
for radiation reaction, cf.~\cite[eq.~after (1.9)]{Nteil}.
\medskip

\noindent
(b) In the context of general relativity, one has to use
the quadrupole moment
\[ Q_{ij}(t)=\int \Big(x_ix_j-\frac{1}{3}|x|^2\delta_{ij}\Big)\rho (t, x)\,dx \]
instead of the dipole moment $D(t)$ and it is the fifth time derivative
which occurs instead of the third before reduction \cite{burke}. This
leads to considerable complications.
\medskip

\noindent
(c) Notice that $D^{[2]}(t)\equiv 0$ for e.g.~spherically
symmetric solutions, whence there is no radiation damping in this case.
}
\end{remark}

It is the purpose of this paper to analyze rigorously
long-time properties of classical solutions to (VPD).
Therefore we first have to deal with the question of global
existence of solutions, e.g.~for smooth data functions
$f^{\pm}(t=0)=f^{\pm}_0$ of compact support, i.e.~such that
\begin{equation}\label{data-f}
   f^{\pm}_0\in C_0^\infty(\R^3\times\R^3),
   \quad f^{\pm}_0\ge 0,\quad\mbox{and}\quad
   f^{\pm}_0(x, v)=0\quad\mbox{for}\quad |x|\ge r_0\,\,\,\mbox{or}
   \,\,\,|v|\ge r_0,
\end{equation}
with some fixed $r_0>0$. Since global existence is a quite
non-trivial issue for Vlasov-Poisson like systems, we provide
a complete existence proof for (VPD) in section \ref{ex-uni-sect}, 
deriving estimates on higher velocity moments
of $f^{\pm}$ along the lines of \cite{lio-pert}.
This approach has been successfully applied to other
related problems as well; cf.~\cite{andre,boc}.

In this manner we obtain

\begin{theorem}\label{glob-ex} If $f^{\pm}_0$ satisfy (\ref{data-f}),
then there is a unique solution 
\[ f^{\pm}\in C^1([0, \infty[\times\R^3\times\R^3) \]
of (VPD) with data $f^{\pm}(t=0)=f^{\pm}_0$.
\end{theorem}

Having ensured that suitable solutions do exist,
we now turn to the decay estimates for quantities related to (VPD).
We define the total energy
\begin{equation}\label{etot-def}
   {\cal E}(t)={\cal E}_{{\rm kin}}(t)+{\cal E}_{{\rm pot}}(t),
\end{equation}
with
\begin{eqnarray}
   {\cal E}_{{\rm kin}}(t) & =& \frac{1}{2}\int\int |v|^2 (f^+(t, x, v)
   +f^-(t, x, v))\,dx dv,\quad\mbox{and} \nonumber \\
   {\cal E}_{{\rm pot}}(t)
   & =& \frac{1}{8\pi}\int |\nabla U(t, x)|^2\,dx,
\label{ekin-def}
\end{eqnarray}
denoting kinetic and potential energy, respectively.

\begin{theorem}\label{dec-summa}
Assume $f^{\pm}_0$ satisfy (\ref{data-f}). Then
\begin{equation}\label{energ-bd}
   \dot{{\cal E}}(t) = -\eps\,{|D^{[2]}(t)|}^2.
\end{equation}
Moreover, the following estimates hold for $t\in [0, \infty[$.
\begin{itemize}
\item[(a)] $\displaystyle {\|\rho^{\pm}(t)\|}_{p; x}
\le C{(1+t)}^{-\frac{3(p-1)}{2p}}$ for $p\in [1, \frac{5}{3}]$;
\item[(b)] $\displaystyle {\|\nabla U(t)\|}_{p; x}
\le C{(1+t)}^{-\frac{5p-3}{7p}}$ for $p\in [2, \frac{15}{4}]$;
\item[(c)] $|D^{[2]}(t)|\le C{(1+t)}^{-\frac{8}{7}}$.
\end{itemize}
In particular, (VPD) does not admit nontrivial
static solutions, and the kinetic energy satisfies
${\cal E}_{{\rm kin}}(t)\to {\cal E}_{\infty}$ as $t\to\infty$
for some ${\cal E}_{\infty}\ge 0$. Moreover, if ${\cal E}(0)>0$ and $\eps>0$
is small enough, then ${\cal E}_\infty>0$.
\end{theorem}

See Section \ref{dec-summa-prf} for the proof. We note that in theorem
\ref{dec-summa} a slow dissipation of energy takes place due to the
``damping term'' $D^{[2]}(t)$, as can be seen from equation
(\ref{energ-bd}).

\begin{remark}\label{bigsav}{\em As an aside, we include a comment
on a relation to the usual Vlasov-Poisson system.
We start with the characteristic equations (\ref{char6}),
(\ref{char5}), i.e.
\begin{eqnarray*}
   \dot{X}^+ & = & V^+ +\eps D^{[2]}(t),\quad
   \dot{V}^+=\nabla U(t, X^+), \\[1ex]
   \dot{X}^- & = & V^- -\eps D^{[2]}(t),
   \quad \dot{V}^-=-\nabla U(t, X^-).
\end{eqnarray*}
Define
\begin{eqnarray*}
   & & \bar{X}^+ = X^+,\quad \bar{V}^+ = V^+ +\eps D^{[2]}(t), \\
   & & \bar{X}^- = X^-,\quad \bar{V}^- = V^- -\eps D^{[2]}(t).
\end{eqnarray*}
Then
\begin{eqnarray*}
   & & \dot{\bar{X}\,}^+ = \,\bar{V}^+,\quad
   \dot{\bar{V}}^+ = \nabla U(t, \bar{X}^+)+\eps\dot{D}^{[2]}(t)
   =\nabla W(t, \bar{X}^+), \\[1ex]
   & & \dot{\bar{X}\,}^- = \,\bar{V}^-,\quad
   \dot{\bar{V}}^- = -\nabla U(t, \bar{X}^-)-\eps\dot{D}^{[2]}(t)
   = -\nabla W(t, \bar{X}^-),
\end{eqnarray*}
where
\[ W(t, x)=U(t, x)+\eps\dot D^{[2]}(t)\cdot x. \]
Also $\Delta W=\Delta U$. Thus we obtain a solution
of the Vlasov-Poisson system where the potential $W$ does not satisfy the
usual boundary conditions. This is similar to the cosmological solutions
of the Vlasov-Poisson system constructed in \cite{rein-rendall}. They are
obtained directly as solutions of a transformed system but are in the end
solutions of the Vlasov-Poisson system with unconventional boundary
conditions. This reformulation gives a simple way of seeing the volume
preserving property of the flow for (VPD), since we know it for
Vlasov-Poisson. It is, however, not hard to see it directly.
}
\end{remark}

\bigskip

\noindent
{\bf Notation.} Throughout the paper, $C$ denotes a general
constant which may change from line to line and which only
depends on $f^{\pm}_0$. If we consider a solution on a fixed
time interval $[0, T]$, and if $C$ additionally depends on $T$,
this is indicated by $C_T$. The usual $L^p$-norm of a function
$\varphi=\varphi(t, x)$ over $x\in\R^3$ is denoted by
${\|\varphi(t)\|}_{p; x}$, and if $\varphi=\varphi(t, x, v)$
and the integrals are to be extended over $(x, v)\in\R^3\times\R^3$,
then we write ${\|\varphi(t)\|}_{p; xv}$. To simplify notation,
an integral $\int$ always means $\int_{\R^3}$.

\bigskip

\noindent
{\bf Acknowledgements.} We wish to thank Thibault Damour, Gerhard Rein,
Gerhard Sch\"afer and Herbert Spohn for discussions and helpful advice.
MK acknowledges support through a Heisenberg fellowship of DFG.


\section{Proof of theorem \ref{dec-summa}}
\label{dec-summa-prf}

We split the proof into several subsections.

\subsection{Energy dissipation}
\label{energ-diss-sect}

We verify (\ref{energ-bd}) and calculate the change of the total energy
${\cal E}(t)$ from (\ref{etot-def}). Due to (\ref{vlas+}) and (\ref{vlas-})
we have
\begin{eqnarray}\label{ekindot}
   \dot{{\cal E}}_{{\rm kin}}(t) & = &
   \frac{1}{2}\int\int v^2 (\dt f^+ +\dt f^-)\,dx dv
   \nonumber \\ & = & \frac{1}{2}\int\int v^2
   \Big(-[v+\eps D^{[2]}(t)]\cdot\dx f^+ -\nabla U\cdot\dv f^+
   \nonumber \\ & & \hspace{5em}
   -[v-\eps D^{[2]}(t)]\cdot\dx f^- +\nabla U\cdot\dv f^-\Big)\,dx dv
   \nonumber \\ & = & \int\int (v\cdot\nabla U)(f^+-f^-)\,dx dv
   =\int\nabla U\cdot j\,dx,
\end{eqnarray}
where
\[ j(t, x)=j^+(t, x)-j^-(t, x),\quad j^{\pm}(t, x)
   =\int vf^{\pm}(t, x, v)\,dv, \]
is the current. The evaluation of $\dot{{\cal E}}_{{\rm pot}}(t)$
is a little more tedious, and for this purpose we will use
\begin{equation}
   {\cal E}_{{\rm pot}}(t)=\frac{1}{8\pi}\int dx\,|\nabla U|^2
   =-\frac{1}{8\pi}\int dx (\Delta U)U
   =\frac{1}{2}\int dx\,(\rho^- -\rho^+)U, \label{epot-def}
\end{equation}
and moreover the representations of the electric field
by means of Coulomb potentials
\begin{eqnarray}
   U(t, x) & = & -\int\frac{dy}{|x-y|}(\rho^+(t, y)-\rho^-(t, y)),
   \nonumber\\ E(t, x) & := & \nabla U(t, x)
   = -\int dy\,\nabla_x\frac{1}{|x-y|}
   (\rho^+(t, y)-\rho^-(t, y)) \nonumber
   \\ & = & \int dy\,\frac{(x-y)}{|x-y|^3}
   (\rho^+(t, y)-\rho^-(t, y)). \label{E-def}
\end{eqnarray}
Then we obtain through the change of variables
$x\leftrightarrow y$ and $v\leftrightarrow w$
\begin{eqnarray*}
   \lefteqn{\dot{{\cal E}}_{{\rm pot}}(t)} \\ & = &
   \frac{1}{2}\int\int dx dv\,\Big\{(\dt\rho^- -\dt\rho^+)U
   +(\rho^- -\rho^+)(\dt U)\Big\} \\
   & = & -\frac{1}{2}\int\int\int\int dx dy dv dw\,
   \Big\{(\dt f^-(t, x, v) -\dt f^+(t, x, v))
   \\ & & \hspace{14em} \times\frac{1}{|x-y|}
(f^+(t, y, w) -f^-(t, y, w))
   \\ & & \hspace{12em} +\,(f^-(t, x, v) -f^+(t, x, v))
   \\ & & \hspace{14em} \times\frac{1}{|x-y|}
   (\dt f^+(t, y, w) -\dt f^-(t, y, w))\Big\} \\ & = &
   -\,\int\int\int\int dx dy dv dw\,
   (\dt f^-(t, x, v) -\dt f^+(t, x, v))\\ & & \hspace{12em}
   \times\frac{1}{|x-y|}(f^+(t, y, w) -f^-(t, y, w)) 
   \\ & = &
\int\int\int\int\frac{dx dy}{|x-y|}\,dv dw\,
   \Big\{[v-\eps D^{[2]}(t)]\cdot\dx f^-(t, x, v)
   -\nabla U(t, x)\cdot\dv f^-(t, x, v)
   \\ & & \hspace{8em}
   -[v+\eps D^{[2]}(t)]\cdot\dx f^+(t, x, v)
   -\nabla U(t, x)\cdot\dv f^+(t, x, v)\Big\}
   \\ & & \hspace{6em}\times(f^+(t, y, w) -f^-(t, y, w))
   \\[1ex] & = & \int\int\int\frac{dx dy}{|x-y|}\,dv\,
   \Big\{[v-\eps D^{[2]}(t)]\cdot\dx f^-(t, x, v)
   -[v+\eps D^{[2]}(t)]\cdot\dx f^+(t, x, v)\Big\}
   \\ & & \hspace{9em}\times(\rho^+(t, y) -\rho^-(t, y))
   \\[1ex] & = & -\,\int\int\int dx dy dv\,
   \bigg(\nabla_x\frac{1}{|x-y|}\bigg)
   \cdot\Big\{[v-\eps D^{[2]}(t)] f^-(t, x, v)
   \\ & & \hspace{9em} 
   -\,[v+\eps D^{[2]}(t)] f^+(t, x, v)\Big\}(\rho^+(t, y) -\rho^-(t, y))
   \\[1ex] & = & \int\int dx dv\,
   \nabla U(t, x)\cdot\Big\{[v-\eps D^{[2]}(t)] f^-(t, x, v)
   -[v+\eps D^{[2]}(t)] f^+(t, x, v)\Big\}
   \\[1ex] & = & \int\int dx dv\,(\nabla U\cdot v)(f^- -f^+)
   +\int\int dx dv\,\nabla U\cdot (-\eps D^{[2]}(t)f^- -\eps D^{[2]}(t)f^+)
   \\[1ex] & = & -\int\nabla U\cdot j\,dx-\eps\,{|D^{[2]}(t)|}^2,
\end{eqnarray*}
recall the definition of $D^{[2]}(t)$ from (\ref{d2def}).
Combining this with (\ref{ekindot}), we see that (\ref{energ-bd}) holds.

\subsection{Decay of the potential energy}

Here we show a $t^{-1}$-decay of the potential
energy ${\cal E}_{{\rm pot}}(t)$ from (\ref{epot-def}).
The result is similar to \cite{illner-rein,perthame}, but the proof requires
appropriate modifications due to the presence of the term $D^{[2]}(t)$.

\begin{lemma}\label{epot-dec} We have
\[ {\cal E}_{{\rm pot}}(t)\le C(1+t)^{-1}\quad\mbox{and}\quad
   \int\int (x-vt)^2 (f^+ +f^-)\,dx dv\le Ct,\quad t\in [0, \infty[, \]
the constants being independent of $\eps\in [0, 1]$.
\end{lemma}
{\bf Proof\,:} Denote
\begin{equation}\label{Rg-def}
   R(t)=\int\int (x-vt)^2 (f^+ +f^-)\,dx dv\quad\mbox{and}\quad
   g(t)=\frac{t^2}{4\pi}\int |\nabla U|^2\,dx=2t^2 {\cal E}_{{\rm pot}}(t).
\end{equation}
Then a short calculation reveals
\begin{eqnarray}\label{dotR}
   \dot{R}(t) & = & -\,2t\,\int (x\cdot\nabla U)\rho\,dx
   +2t^2\,\int \nabla U\cdot j\,dx
\nonumber \\ & & 
   +\,2\eps D^{[2]}(t)\cdot\,\bigg(\int\int (x-tv)(f^+ -f^-)\,dxdv\bigg).
\end{eqnarray}
Inserting (\ref{E-def}) for $\nabla U$ and with $x\cdot (x-y)\,|x-y|^{-3}
=|x-y|^{-1}+y\cdot (x-y)\,|x-y|^{-3}$, we see that
\begin{equation}\label{eq1}
   \int (x\cdot\nabla U)\rho\,dx=-\frac{1}{2}\,\int U\rho\,dx
   =\frac{1}{8\pi}\,\int {|\nabla U|}^2\,dx.
\end{equation}
On the other hand, (\ref{ekindot}) and the energy identity
(\ref{energ-bd}) imply
\begin{equation}\label{eq2}
   \int\nabla U\cdot j\,dx=\dot{{\cal E}}_{{\rm kin}}(t)
   =-\dot{{\cal E}}_{{\rm pot}}(t)-\eps\,{|D^{[2]}(t)|}^2
   =-\,\frac{1}{8\pi}\,\frac{d}{dt}\,\int {|\nabla U|}^2\,dx
   -\eps\,{|D^{[2]}(t)|}^2.
\end{equation}
Using (\ref{eq1}) and (\ref{eq2}) in (\ref{dotR}), it follows that
\begin{eqnarray*}
   \dot{R}(t) & = & -\,\frac{t}{4\pi}\,\int {|\nabla U|}^2\,dx
   +2t^2\,\bigg(-\,\frac{1}{8\pi}\,\frac{d}{dt}\,\int {|\nabla U|}^2\,dx
   -\eps\,{|D^{[2]}(t)|}^2\bigg)
   \\ & & +\,2\eps D^{[2]}(t)\cdot\,
   \bigg(\int\int (x-tv)(f^+ -f^-)\,dxdv\bigg).
\end{eqnarray*}
By means of $g$ from (\ref{Rg-def}), this may be rewritten as
\begin{eqnarray}\label{dot-Rg}
   \frac{d}{dt}\,\Big(R(t)+g(t)\Big)
   & = & \frac{g(t)}{t}-2\eps t^2\,{|D^{[2]}(t)|}^2
\nonumber \\ & & 
   +\,2\eps D^{[2]}(t)\cdot\,\bigg(\int\int (x-tv)(f^+ -f^-)\,dxdv\bigg).
\end{eqnarray}
Compared to \cite[p.~1412]{illner-rein}, it is now necessary to
see how the two terms with $D^{[2]}(t)$ contribute.

First we consider the case that $t>0$ is such that
\[ t^2\,|D^{[2]}(t)|\le\bigg|\int\int (x-tv)(f^+ -f^-)\,dxdv\bigg|. \]
Then we obtain from (\ref{dot-Rg}) that
\begin{eqnarray}\label{Rg-esti}
   \frac{d}{dt}\,\Big(R(t)+g(t)\Big)
   & \le & \frac{g(t)}{t}
   +2\eps |D^{[2]}(t)|\,\bigg|\int\int (x-tv)(f^+ -f^-)\,dxdv\bigg|
   \nonumber \\ & \le & \frac{g(t)}{t}
   +2\eps t^{-2}\,{\bigg|\int\int (x-tv)(f^+ -f^-)\,dxdv\bigg|}^2.
\end{eqnarray}
However, if
\[ t^2\,|D^{[2]}(t)|\ge \bigg|\int\int (x-tv)(f^+ -f^-)\,dxdv\bigg|, \]
then (\ref{dot-Rg}) yields
\[ \frac{d}{dt}\,\Big(R(t)+g(t)\Big)\le\frac{g(t)}{t}, \]
hence (\ref{Rg-esti}) is verified for all $t>0$. In order to have
bounds below independent of, say, $\eps\in [0, 1]$, we
modify (\ref{Rg-esti}) to
\begin{eqnarray}\label{Rg-esti-2}
   \frac{d}{dt}\,\Big(R(t)+g(t)\Big)\le\frac{g(t)}{t}
   +2 t^{-2}\,{\bigg|\int\int (x-tv)(f^+ -f^-)\,dxdv\bigg|}^2.
\end{eqnarray}
To further exploit this,
we next note that due to H\"ol\-der's inequality and by lemma \ref{bd-1}
below with $p=0$
\begin{eqnarray}\label{eq5}
   \lefteqn{{\bigg|\int\int (x-tv)(f^+ -f^-)\,dxdv\bigg|}^2} \nonumber \\
   & \le & \bigg(\int\int (x-tv)^2[f^+ +f^-]\,dxdv\bigg)
   \,\bigg(\int\int [f^+ +f^-]\,dxdv\bigg)\le CR(t),
\end{eqnarray}
thus by (\ref{Rg-esti-2})
\[ \frac{d}{dt}\,\Big(R(t)+g(t)\Big)\le\frac{g(t)}{t}+Ct^{-2}R(t),
   \quad t>0. \]
Integrating over $t\in [1, T]$, we see that
\begin{equation}\label{eq3}
   R(T)\le R(T)+g(T)\le C+\int_1^T \frac{g(t)}{t}\,dt
   +C\,\int_1^T t^{-2}R(t)\,dt,\quad T\ge 1.
\end{equation}
Therefore
\begin{equation}\label{R-absch}
   R(T)\le C\bigg(1+\int_1^T \frac{g(t)}{t}\,dt\bigg),\quad T\ge 1,
\end{equation}
by Gronwall's lemma. Using this in (\ref{eq3}), we find
\begin{eqnarray*}
   g(T) & \le & C+\int_1^T \frac{g(t)}{t}\,dt
   +C\,\int_1^T\bigg(1+\int_1^t \frac{g(s)}{s}\,ds\bigg)
   \,\frac{dt}{t^2} \\
   & \le & C+\int_1^T \frac{g(t)}{t}\,dt
   +\,C\int_1^T\bigg(\frac{1}{t}-\frac{1}{T}\bigg)\,\frac{g(t)}{t}\,dt,
\end{eqnarray*}
and consequently
\[ g(T)\le CT,\quad T\ge 1, \]
again by Gronwall's lemma. According to the definition of $g$,
this proves the $t^{-1}$-decay of ${\cal E}_{{\rm pot}}(t)$,
and then (\ref{R-absch}) shows that $R(T)\le CT$ holds as well.
This completes the proof of lemma \ref{epot-dec}. {\hfill$\Box$}\bigskip

\subsection{Some general estimates}

We digress now from the proof of theorem \ref{dec-summa} and note some
useful estimates that will also play a role later for the global existence
of solutions, cf.~theorem \ref{glob-ex}. For $p\in [0, \infty[$ 
define the velocity moments
\begin{equation}\label{mom-def}
   M^{\pm}_p(t)=\int\int |v|^p f^{\pm}(t, x, v)\,dx dv,
   \quad\mbox{and}\quad M_p(t)=\sup_{s\in [0, t]}\Big(M^+_p(s)
   +M^-_p(s)\Big). 
\end{equation}

\begin{lemma}\label{bd-1} For $t\in [0, \infty[$ we have
\[ {\|f^{\pm}(t)\|}_{\infty;\,xv}\le C\quad\mbox{and}\quad
   M_p(t)\le C,\quad p\in [0, 2]. \]
\end{lemma}
{\bf Proof\,:} For fixed $t\in [0, \infty[$ let
$({\cal X}(s), {\cal V}(s))=({\cal X}(s; t, x, v), {\cal V}(s; t, x, v))$
denote the characteristics from (\ref{char6}) associated with (\ref{vlas+}), i.e.
\begin{equation}\label{doz}
   \left(\begin{array}{c} \dot{{\cal X}}(s)
   \\ \dot{{\cal V}}(s)\end{array}\right)
   =\left(\begin{array}{c} {\cal V}(s)+\eps D^{[2]}(s)
   \\ \nabla U(s, {\cal X}(s))
   \end{array}\right), \quad
   \left(\begin{array}{c} {\cal X}(t) \\ {\cal V}(t)\end{array}\right)
   =\left(\begin{array}{c} x \\ v\end{array}\right).
\end{equation}
Then $\frac{\partial}{\partial s}[f^+(s, {\cal X}(s), {\cal V}(s))]=0$
shows $f^+(t, x, v)=f^+_0({\cal X}(0), {\cal V}(0))$,
and hence the first bound follows.
Concerning the second, 
\[ M_2(t)=2\sup_{s\in [0, t]}{\cal E}_{{\rm kin}}(s)
   \le 2\sup_{s\in [0, t]}{\cal E}(s)=2 {\cal E}(0) \] 
by (\ref{energ-bd}). Also $M_0^{\pm}(t)=M_0^{\pm}(0)$, as $(x, v)\mapsto
({\cal X}(0; t, x, v), {\cal V}(0; t, x, v))$ is a volume-preserving
diffeomorphism of $\R^3\times\R^3$, due to the fact that the right-hand side
of the ODE in (\ref{doz}) has divergence
${\rm div}={\rm div}_{({\cal X}, {\cal V})}$ zero;
see also lemma \ref{trafo} below and remark \ref{bigsav}.
Observing that ${|v|}^p\le 1+{|v|}^2$
for $v\in\R^3$ and $p\in [0, 2]$ completes the proof. {\hfill$\Box$}\bigskip

\begin{lemma}\label{R-arg} Let $f=f(x, v)\in L^\infty(\R^3\times\R^3)$ 
be a nonnegative function such that $\int\int |v|^p f(x, v)\,dxdv<\infty$ 
for some $p\in [0, \infty[$, and define $\phi(x)=\int f(x, v)\,dv$, $x\in\R^3$. 
Then 
\begin{equation}\label{pilat}
   {\|\phi\|}_{\frac{3+p}{3};\,x}\le C\,{\|f\|}_{\infty;\,xv}^{\frac{p}{3+p}}
   \,{\bigg(\int\int |v|^p f(x, v)\,dxdv\bigg)}^{\frac{3}{3+p}}.
\end{equation}
Here $C$ depends only on $p$.
\end{lemma}
{\bf Proof\,:} The argument is well-known, but indicated for
completeness. We split
\[ \phi(x)\le\int_{|v|\le R}f(x, v)\,dv+\int_{|v|\ge R} f(x, v)\,dv
   \le \frac{4\pi}{3}R^3\,{\|f\|}_{\infty;\,xv}
   +R^{-p}\int |v|^p f(x, v)\,dv \]
and optimize in $R$ to find
\[ \phi(x)\le C\,{\|f\|}_{\infty;\,xv}^{\frac{p}{3+p}}
   \,{\bigg(\int |v|^p f(x, v)\,dv\bigg)}^{\frac{3}{3+p}}, \]
whence integration w.r.t.~$x$ yields (\ref{pilat}).
{\hfill$\Box$}\bigskip

\begin{lemma}\label{falt-lem} We have
\[ {\bigg\|\bigg(\nabla\frac{1}{|x|}\bigg)\ast\rho\,\bigg\|}_{q;\,x}
   \le C {\|\rho\|}_{p;\,x},\quad q\in ]\frac{3}{2}, \infty[,
   \quad p=\frac{3q}{3+q}. \]
In addition,
\[ {\bigg\|\bigg(\nabla\frac{1}{|x|}\bigg)\ast
   {\rm div}\,\Gamma\,\bigg\|}_{q;\,x}
   \le C {\|\Gamma\|}_{q;\,x},\quad q\in ]1, \infty[, \]
for smooth and compactly supported vector fields $\Gamma: \R^3\to\R^3$.
\end{lemma}
{\bf Proof\,:} The first estimate is a consequence of the classical
Hardy-Littlewood-Sobolev inequality; see \cite[Thm.~4.5.3]{horm}.
Concerning the second, we note that integration by parts reveals
\begin{equation}\label{laubr}
   \int\frac{x-y}{|x-y|^3}\,{\rm div}\,\Gamma(y)\,dy
   =\frac{4\pi}{3}\,\Gamma(x)-\lim_{\eps\to 0}
   \,\int_{|x-y|\ge\eps}\Gamma(x-y)\cdot g(y)\,dy,
\end{equation}
with $g(y)=\frac{1}{|y|^3}\,G(y)$, where $G(y)=(-{\rm Id})+\frac{3}{|y|^2}(y\otimes y)
\in\R^{3\times 3}$. Since $G$ is bounded in $\R^3\setminus\{0\}$, homogeneous
of degree zero, and satisfies $\int_{|y|=1}G(y)\,d^2y=0$, the Cal\-der\-\'{o}n-Zygmund
inequality \cite[Thm.~4.31]{adams} implies that the second term on the right-hand side
of (\ref{laubr}) defines a bounded operator $L^q(\R^3)\to L^q(\R^3)$; in view of the
compact support of the $\Gamma$'s it is not necessary that $G$ has compact support.
{\hfill$\Box$}\bigskip

\begin{lemma} For $t\in [0, \infty[$ we have
\begin{equation}\label{terrac}
   {\|\rho^{\pm}(t)\|}_{p;\,x}\le
   C{M_{3(p-1)}(t)}^{\frac{1}{p}},\quad p\in [1, \infty[,
\end{equation}
as well as
\begin{equation}\label{gdby}
   {\|\nabla U(t)\|}_{q;\,x}\le C {\|\rho(t)\|}_{\frac{3q}{3+q};\,x}
   \le C{M_{\frac{6q-9}{3+q}}(t)}^{\frac{3+q}{3q}},\quad
   q\in ]\frac{3}{2}, \infty[.
\end{equation}
Moreover,
\begin{equation}\label{justiro}
   |D^{[2]}(t)|\le {\|\nabla U(t)\|}_{p;\,x}
   \,{\|\rho(t)\|}_{p';\,x},\quad p\in [1, \infty].
\end{equation}
\end{lemma}
{\bf Proof\,:} According to lemma \ref{R-arg} and lemma \ref{bd-1},
\[ {\|\rho^{\pm}(t)\|}_{\frac{3+\alpha}{3};\,x}
   \le C\,{\|f^{\pm}(t)\|}_{\infty;\,xv}^{\frac{\alpha}{3+\alpha}}
   \,{M_\alpha(t)}^{\frac{3}{3+\alpha}}
   \le C {M_\alpha(t)}^{\frac{3}{3+\alpha}} \]
for all $\alpha\ge 0$, hence (\ref{terrac}) holds.
Due to (\ref{E-def}) we see lemma \ref{falt-lem} applies
to yield, for $q\in ]\frac{3}{2}, \infty[$ and with $p=\frac{3q}{3+q}$,
together with (\ref{terrac})
\[ {\|\nabla U(t)\|}_{q;\,x}\le C\Big(
   {\|\rho^+(t)\|}_{p;\,x}+{\|\rho^-(t)\|}_{p;\,x}\Big)
   \le C{M_{3(p-1)}(t)}^{\frac{1}{p}}. \]
Expressing $p$ through $q$, we arrive at (\ref{gdby}).
The estimate on $|D^{[2]}(t)|$ is a consequence of
(\ref{d2def}) and H\"older's inequality. {\hfill$\Box$}\bigskip

\subsection{Proof of theorem \ref{dec-summa} (completed)}

{}From lemma \ref{epot-dec} we additionally obtain,
analogously to \cite{illner-rein},
the following information.

\begin{cor}\label{oldpi} Under the assumptions of theorem \ref{dec-summa},
we moreover have
\begin{equation}\label{fnch1}
   {\|\rho^{\pm}(t)\|}_{\frac{5}{3};\,x}\le C (1+t)^{-3/5},
   \quad t\in [0, \infty[,
\end{equation}
and
\begin{equation}\label{fnch2}
   {\|\nabla U(t)\|}_{\frac{15}{4};\,x}\le C (1+t)^{-3/5},
   \quad t\in [0, \infty[.
\end{equation}
\end{cor}
{\bf Proof\,:} Using lemma \ref{bd-1} and lemma \ref{epot-dec}
we can split
\begin{eqnarray*}
   \rho^{\pm}(t, x) & \le & \int_{\{v: |x-tv|\le R\}}
   f^{\pm}(t, x, v)\,dv+R^{-2}\,\int_{\{v: |x-tv|\ge R\}}
   (x-tv)^2\,f^{\pm}(t, x, v)\,dv
   \\ & \le & C R^3 t^{-3}+R^{-2}\,\int (x-tv)^2\,(f^+ +f^-) (t, x, v)\,dv
   \le C R^3 t^{-3}+CR^{-2}t,
\end{eqnarray*}
and then choose the optimal $R\cong t^{4/5}$ to obtain (\ref{fnch1}).
Concerning (\ref{fnch2}), this follows from (\ref{fnch1})
and the first inequality in (\ref{gdby}) with $q=\frac{15}{4}$. {\hfill$\Box$}\bigskip

\begin{lemma}\label{a-c} Assertions (a)--(c) of theorem \ref{dec-summa}
are satisfied.
\end{lemma}
{\bf Proof\,:} {}From lemma \ref{bd-1} and corollary \ref{oldpi}
we know that ${\|\rho^{\pm}(t)\|}_{1;\,x}\le C$, and moreover
${\|\rho^{\pm}(t)\|}_{\frac{5}{3};\,x}\le C (1+t)^{-3/5}$.
We may also estimate ${\|\nabla U(t)\|}_{2;\,x}=\sqrt{8\pi}
{\cal E}_{{\rm pot}}(t)^{1/2}\le C (1+t)^{-1/2}$
as well as ${\|\nabla U(t)\|}_{\frac{15}{4};\,x}\le C (1+t)^{-3/5}$
by lemma \ref{epot-dec} and corollary \ref{oldpi}.
Hence the general interpolation estimate
\[ {\|\phi\|}_p\le {\|\phi\|}_{q_1}^{\alpha}\,
   {\|\phi\|}_{q_2}^{1-\alpha},\quad p\in [q_1, q_2],
   \quad\frac{1}{p}=\frac{\alpha}{q_1}+\frac{1-\alpha}{q_2}, \]
yields (a) and (b). For (c), we use (\ref{justiro}) with
$p=5/2$ and $p'=5/3$, (a), and (b) to see that
\begin{equation}\label{wotuzo}
   |D^{[2]}(t)|\le {\|\nabla U(t)\|}_{\frac{5}{2};\,x}
   \,{\|\rho(t)\|}_{\frac{5}{3};\,x}
   \le C{(1+t)}^{-19/35}(1+t)^{-3/5}=C{(1+t)}^{-8/7},
\end{equation}
as was to be shown. {\hfill$\Box$}\bigskip

\begin{remark}{\rm The estimates derived thus far suggest
that the optimal decay rate for $D^{[2]}(t)$ be $|D^{[2]}(t)|\sim t^{-3/2}$
rather than $|D^{[2]}(t)|\sim t^{-8/7}$, for the following reason:
by H\"older's inequality and lemma \ref{epot-dec} the quantity 
$I(t)=\int\int (x-vt)(f^+ -f^-)\,dxdv$ satisfies
\[ |I(t)|\le C\bigg(\int\int (x-vt)^2(f^+ +f^-)\,dxdv\bigg)^{1/2}\le C(1+t)^{1/2}, \]
thus we might expect $\dot{I}(t)\sim t^{-1/2}$. On the other hand,
direct calculation shows
\[ \dot{I}(t)=\bigg(\eps\int\int (f^+ +f^-)\,dxdv-t\bigg)D^{[2]}(t)\sim
   (-t)D^{[2]}(t), \]
whence we should have $|D^{[2]}(t)|\sim t^{-3/2}$. This decay
would also be obtained if it were possible to use
theorem \ref{dec-summa}(a) and (b) with $p=\frac{15}{4}$ and
$p=\frac{15}{11}$, respectively, since then (\ref{justiro}) would yield
\begin{eqnarray*} 
   |D^{[2]}(t)| & \le & {\|\nabla U(t)\|}_{\frac{15}{11};\,x}
   \,{\|\rho(t)\|}_{\frac{15}{4};\,x}
   \\ & \le & C{(1+t)}^{-\frac{5(15/11)-3}{7(15/11)}}
   {(1+t)}^{-\frac{3((15/4)-1)}{2(15/4)}}=C{(1+t)}^{-3/2}.
\end{eqnarray*} 
However, the necessary decay estimates for such $p$-norms of $\nabla U(t)$
and $\rho(t)$ could not be proved.
}
\end{remark}\bigskip

\begin{cor}\label{no-static}
There are no nontrivial static solutions of (VPD), and
${\cal E}_{{\rm kin}}(t)\to {\cal E}_{\infty}\ge 0$
as $t\to\infty$. If ${\cal E}(0)>0$ and $\eps>0$ is sufficiently small,
then ${\cal E}_\infty>0$.
\end{cor}
{\bf Proof\,:} If (VPD) had a static solution $f^{\pm}(t)\equiv f^{\pm}_0$,
then ${\cal E}_{{\rm pot}}(t)\equiv 0$, whence $\nabla U=0$.
This in turn yields $D^{[2]}(t)\equiv 0$ by definition.
Consequently, the Vlasov equations (\ref{vlas+}), (\ref{vlas-})
reduce to $\dt f^{\pm} + v\cdot\dx f^{\pm}=0$ with unique solution
$f^{\pm}(x, v)=f^{\pm}_0(x-vt, v)$. But then
we see $\rho^{\pm}(x)=\int f^{\pm}_0(x-vt, v)\,dv
=t^{-3}\int f^{\pm}_0(w, t^{-1}[x-w])\,dw$,
showing as $t\to\infty$ that the solution has to be trivial.
To prove the assertion concerning ${\cal E}_{{\rm kin}}(t)$,
note that, since ${\cal E}(t)$ is decaying by (\ref{energ-bd}),
${\cal E}(t)\to {\cal E}_{\infty}\ge 0$ as $t\to\infty$.
But ${\cal E}(t)={\cal E}_{{\rm kin}}(t)+{\cal E}_{{\rm pot}}(t)$ and
${\cal E}_{{\rm pot}}(t)\to 0$, hence the first claim follows.
For the second, denote $C_1$ the constant on the right-hand side
of (\ref{wotuzo}). Since all bounds are derived from lemma \ref{epot-dec},
we note that $C_1$ is independent of $\eps\in [0, 1]$.
Integrating (\ref{energ-bd}) yields
\[ {\cal E}_{{\rm kin}}(t)+{\cal E}_{{\rm pot}}(t)
   ={\cal E}(0)-\eps\int_0^t |D^{[2]}(s)|^2\,ds, \]
thus as $t\to\infty$, provided that ${\cal E}_\infty=0$,
by (\ref{wotuzo})
\[ {\cal E}(0)=\eps\int_0^\infty |D^{[2]}(s)|^2\,ds
   \le C_1^2\eps\,\int_0^\infty {(1+s)}^{-16/7}\,ds
   =(7C_1^2/9)\eps. \]
So if we choose $\eps<(9/7C_1^2){\cal E}(0)$,
then necessarily ${\cal E}_\infty>0$. {\hfill$\Box$}\bigskip

Taking into account Section \ref{energ-diss-sect}, lemma \ref{a-c},
and corollary \ref{no-static}, we note that the proof
of theorem \ref{dec-summa} is complete.

\begin{remark}{\rm With regard to corollary \ref{no-static},
${\cal E}_\infty=\lim_{t\to\infty}{\cal E}_{{\rm kin}}(t)>0$
was to be expected, since otherwise the particle velocities
would have to tend to zero. It is, however, not surprising that
at late times, when we are in a small data regime and the radiation
reaction force is getting small, the solution behaves like a
solution of the Vlasov-Poisson system with small data. In that
case the particles travel with constant non-zero velocity
at late times, as shown in \cite{bar-deg}.}
\end{remark}

We note a further consequence of the foregoing estimates.

\begin{cor} If ${\cal E}_\infty>0$, then
\[ C_1 t-C_2\le\int\int (x\cdot v)(f^+(t, x, v)+f^-(t, x, v))\,dx dv
   \le C_3(1+t),\quad t\in [0, \infty[, \]
for constants $C_1, C_2, C_3>0$.
\end{cor}
{\bf Proof\,:} Denote $S(t)=\int\int (x\cdot v)(f^+ +f^-)\,dx dv$.
In view of lemma \ref{epot-dec} we obtain
\begin{eqnarray*}
   2t^2 {\cal E}_{{\rm kin}}(t) & = & \int\int (x-vt)^2 (f^+ +f^-)\,dx dv
   -\int\int x^2 (f^+ +f^-)\,dx dv+2tS(t) \\
   & \le & C(1+t)+2tS(t),
\end{eqnarray*}
whence $t^2 {\cal E}_\infty\le C(1+t)+2tS(t)$ for $t$ large enough.
To prove the upper bound, note that $Q(t)=\int\int x^2(f^+ +f^-)\,dx dv$
satisfies $\dot{Q}(t)=2S(t)+2\eps D^{[2]}(t)\cdot \int\int x(f^+ -f^-)\,dx dv$,
as follows by a straightforward calculation.
Therefore utilizing H\"older's inequality we obtain
\[ \dot{Q}(t)\le 2Q(t)^{1/2}(2{\cal E}_{{\rm kin}}(t))^{1/2}
   +C(1+t)^{-8/7}Q(t)^{1/2}\le CQ(t)^{1/2}. \]
Consequently, $Q(t)\le C(1+t)^2$, and this in turn yields,
once more by H\"older's inequality, $|S(t)|\le Q(t)^{1/2}
(2{\cal E}_{{\rm kin}}(t))^{1/2}\le C(1+t)$.
{\hfill$\Box$}\bigskip


\section{Existence of solutions}
\label{ex-uni-sect}

As mentioned in the introduction, the proof follows \cite{lio-pert}.
The idea is to decompose the field $E=\nabla U=E_1+F$
in a ``far field'' $F$ that is small, and in some complementary part $E_1$
which is of higher regularity than $E$ itself. (More precisely,
${\|E_1(t)\|}_{p;\,x}\le C$ for every $p\in [1, \frac{15}{4}[$ can be achieved.)
According to this splitting, we write the Vlasov equations
(\ref{vlas+}) and (\ref{vlas-}) in the form
\begin{eqnarray}
   \dt f^+ + (v+\eps D^{[2]}(t))\cdot\dx f^+
   +F \cdot\dv f^+ & = & -E_1\cdot\dv f^+, \label{vlas+1} \\[1ex]
   \dt f^- + (v-\eps D^{[2]}(t))\cdot\dx f^-
   - F\cdot\dv f^- & = & E_1\cdot\dv f^-. \label{vlas-1}
\end{eqnarray}
Since $F$ is small, the characteristics of e.g.~(\ref{vlas+1})
should behave as
\begin{equation}\label{char-appr}
   {\cal X}(s)\approx x+(s-t)v+\eps\int_t^s D^{[2]}(\tau)\,d\tau,
   \quad {\cal V}(s)\approx v,
\end{equation}
which is close to a free streaming, at least in case $D^{[2]}$ were
not present. Writing $\rho^{\pm}(t, x)$ as a suitable integral
over characteristics, it then turns out that in order to derive
the necessary estimates for global existence (on higher moments),
it is possible to use a rigorous form of (\ref{char-appr}).
In particular, one may verify that
\[ \bigg|\det\bigg(\frac{\partial {\cal X}}{\partial v}(s)\bigg)^{-1}\bigg|
   \approx |s-t|^{-3}\quad\mbox{and}\quad
   \bigg|\frac{\partial x}{\partial {\cal V}}(s)\bigg|\approx |s-t|, \]
as is important to transform away the characteristics. The main
point to note here is that the term with $D^{[2]}$ drops if we
take derivatives in (\ref{char-appr}) w.r.t. $x$ or $v$, and
hence the arguments from \cite{lio-pert} can be expected to carry over.
Having derived the higher moment bounds
\[ M_m(t)\le C,\quad t\in [0, T],\quad m\in ]3, \frac{51}{11}[, \]
then a standard argument yields the global existence
of classical solutions for (VPD).

It should finally be remarked that we did not succeed in generalizing
the proofs of global existence for the usual Vlasov-Poisson
system that bound the increase in velocity along
a characteristic; see \cite{pfaff, schaeff, gerh}, and \cite{pulvi}
for a recent application. The reason for this is that, when estimating
the ``ugly'' term, an  $\ddot{{\cal X}}(s)$ will appear, which in our case
will lead to the expression $\dot{D}^{[2]}(s)$ that could not be bounded
well enough to make the proof work.

\subsection{Local existence}

Similar to the case of the usual Vlasov-Poisson system,
cf.~\cite{batt}, where this is contained implicitly,
an iteration scheme may be set up to yield the local existence
of a solution and a criterion when a local solution in fact will be global.

\begin{theorem}\label{loc-ex} Suppose $f^{\pm}_0$ satisfy (\ref{data-f}).
Then there exist unique solutions
\[ f^{\pm}\in C^1([0, T_\ast[\times\R^3\times\R^3) \]
of (\ref{DelU}), (\ref{rho-def}), (\ref{vlas1}), and (\ref{vlas2})
with data $f^{\pm}(t=0)=f^{\pm}_0$, on a maximal time interval
of existence $[0, T_\ast[$. If moreover
\begin{equation}\label{P-def}
   P=P^+ +P^-,\quad\mbox{with}\quad
   P^{\pm}(t)=\sup\Big\{|v|:\,\,\exists\,x\in\R^3\,\,
   \exists\,s\in [0, t]:\,(x, v)\in {\rm supp}f^{\pm}(s)\Big\},
\end{equation}
is bounded on $[0, T_\ast[$, then $T_\ast=\infty$.
\end{theorem}
We will not go into the proof of this result.

\subsection{Some preliminary estimates}

We first need to derive some a priori bounds.
For this we consider a classical solution
of the system that exists for times $t\in [0, T]$.
Note that all estimates from the previous sections remain
valid on any interval where the solution exists.

\begin{lemma}\label{bd-3} For $t\in [0, T]$ we have
\[ M_p(t)\le C_T\Big(1+\sup_{s\in [0, t]}
   {\|\nabla U(s)\|}_{3+p;\,x}^{3+p}\Big),\quad p\in [1, \infty[. \]
In addition,
\begin{equation}\label{mome-mome}
   M_p(t)\le C_T\bigg(1+{M_{3(\frac{3+2p}{6+p})}(t)}
   ^{\frac{6+p}{3}}\bigg),\quad p\in [1, \infty[.
\end{equation}
\end{lemma}
{\bf Proof\,:} Recalling (\ref{mom-def}), from (\ref{vlas+})
and H\"older's inequality it follows that
\begin{eqnarray*}
   \frac{d}{dt}M_p^+(t) & = & \int\int\,|v|^p \Big\{-[v+\eps D^{[2]}(t)]
   \cdot\dx f^+ -\nabla U\cdot\dv f^+\Big\}\,dx dv
   \\ & = & -\int\int\,|v|^p\,\nabla U\cdot\dv f^+\,dx dv
   =p\,\int\int\,|v|^{p-2}(v\cdot\nabla U) f^+\,dx dv
   \\ & \le & C {\|\nabla U(t)\|}_{3+p;\,x}
   \,{\bigg\|\int\,|v|^{p-1}
   f^+(t, \cdot, v)\,dv\bigg\|}_{\frac{3+p}{2+p};\,x}.
\end{eqnarray*}
Now
\[ {\bigg\|\int\,|v|^{p-1}
   f^{\pm}(t, \cdot, v)\,dv\bigg\|}_{\frac{3+p}{2+p};\,x}
   \le C{M_p^{\pm}(t)}^{\frac{2+p}{3+p}} \]
by an argument similar to the proof of lemma \ref{R-arg},
and this yields
\[ \bigg|\frac{d}{dt}M_p^{\pm}(t)\bigg|\le
   C{\|\nabla U(t)\|}_{3+p;\,x}\,{M_p(t)}^{\frac{2+p}{3+p}}. \]
Since $M_p(\cdot)$ is increasing, it is differentiable
a.e.~in $t$, with
\begin{equation}\label{Mp-diffint}
   \frac{d}{dt}M_p(t)\le
   \sup_{s\in [0, t]}\,\bigg|\frac{d}{dt}\Big(M_p^+(s)
   +M_p^-(s)\Big)\bigg|\le
   C\sup_{s\in [0, t]}\,{\|\nabla U(s)\|}_{3+p;\,x}
   \,{M_p(t)}^{\frac{2+p}{3+p}}.
\end{equation}
Integration of this differential inequality gives the claim.
Finally, for (\ref{mome-mome}) we observe that by the first part
and (\ref{gdby}) with $q=3+p$
\[ M_p(t)\le C_T+C_T\sup_{s\in [0, t]}
   {\|\nabla U(s)\|}_{3+p;\,x}^{3+p}
   \le C_T+C_T\,{M_{\frac{6q-9}{3+q}}(t)}^{\frac{3+q}{3q}(3+p)}, \]
and $\frac{6q-9}{3+q}=3(\frac{3+2p}{6+p})$.
{\hfill$\Box$}\bigskip

\subsection{Estimates for higher moments}

For $R>0$ choose a radially symmetric function $\chi_R\in C_0^\infty(\R^3)$
with $\chi_R(x)\in [0, 1]$ for $x\in\R^3$, $\chi_R(x)=1$ for $|x|\le R$,
and $\chi(x)=0$ for $|x|\ge 2R$.
Correspondingly we decompose the electric field $E(t, x)$ from (\ref{E-def}) as
\[ E(t, x)=E_1(t, x)+F(t, x), \]
with
\begin{eqnarray}\label{E1-def}
   E_1(t, x) & = & \int\chi(x-y)\frac{(x-y)}{|x-y|^3}(\rho^+(t, y)-\rho^-(t, y))\,dy
   \nonumber \\ & = & 
   -\,\bigg(\chi\,\nabla\frac{1}{|x|}\bigg)\ast (\rho^+(t)-\rho^-(t))(x).
\end{eqnarray}
Some useful estimates on $E_1$ and $F$ are stated in lemma \ref{e1f-lem}
below. Then we write the Vlasov equations
(\ref{vlas+}) and (\ref{vlas-}) in the form (\ref{vlas+1}) and
(\ref{vlas-1}). This can be used to derive a representation formula for $\rho^{\pm}(t, x)$,
and for simplicity we will consider only $\rho^+(t, x)$.
We fix $x, v\in\R^3$ and $t\in [0, T]$, and denote $(X(s), V(s))=(X(s; x, v),
V(s; x, v))$ for $s\in [0, t]$ the solution of the characteristic system
\begin{equation}\label{char-syst}
   \left(\begin{array}{c} \dot{X}(s) \\ \dot{V}(s)\end{array}\right)
   =\left(\begin{array}{c} -V(s)-\eps D^{[2]}(t-s) \\ -F(t-s, X(s))\end{array}\right),
   \quad
   \left(\begin{array}{c} X(0) \\ V(0)\end{array}\right)
   =\left(\begin{array}{c} x \\ v\end{array}\right),
\end{equation}
associated with (\ref{vlas+1}). Since
\begin{eqnarray*}
   \lefteqn{\frac{\partial}{\partial s}[f^+(t-s, X(s), V(s))]}
   \\ & = & -\dt f^+(t-s, X(s), V(s))-[V(s)+\eps D^{[2]}(t-s)]\cdot
   {\partial_X}f^+(t-s, X(s), V(s))
   \\ & & -F(t-s, X(s))\cdot {\partial_V}f^+(t-s, X(s), V(s))
   \\ & = & E_1(t-s, X(s))\cdot {\partial_V}f^+(t-s, X(s), V(s))
\end{eqnarray*}
by (\ref{vlas+1}), it follows through integrating $\int_0^t ds(\ldots)$
and $\int dv(\ldots)$ that
\begin{eqnarray}
   \rho^+(t, x) & = & \int dv\,f^+_0(X(t), V(t))
   \nonumber  \\& & -\,\int_0^t ds\,\int dv\,E_1(t-s, X(s))
   \cdot {\partial_V}f^+(t-s, X(s), V(s))
   \nonumber \\ & = & \int dv\,f^+_0(X(t), V(t))
   -\int_0^t ds\,\int dv\,{\rm div}_V [E_1 f^+](t-s, X(s), V(s)),\quad
   \label{rho+form}
\end{eqnarray}
where $f^+_0=f^+(t=0)$, and $[E_1 f^+](\tau, X, V)=E_1(\tau, X)f^+(\tau, X, V)$;
note that the dependence on $x$ and $v$ in (\ref{rho+form})
enters via $X(s)$ and $V(s)$. To rewrite (\ref{rho+form})
appropriately, define
\[ G(X, V)=[E_1 f^+](t-s, X, V)\quad\mbox{and}
   \quad\tilde{G}(x, v)=G(X(s; x, v), V(s; x, v)). \]
By lemma \ref{trafo} below we then have $G(X, V)=\tilde{G}(x(s; X, V), v(s; X, V))$,
and consequently
\[ {\rm div}_V G={\rm div}_x\bigg(\frac{\partial x}{\partial V}
   \cdot\tilde{G}\bigg)-{\rm div}_x
   \bigg(\frac{\partial x}{\partial V}\bigg)\cdot\tilde{G}
   + \sum_{i, j=1}^3 \bigg(\frac{\partial \tilde{G}_i}{\partial v_j}\bigg)
   \bigg(\frac{\partial v_j}{\partial V_i}\bigg), \]
where ${\rm div}_x(\frac{\partial x}{\partial V})\cdot\tilde{G}
=\sum_{i, j=1}^3 \tilde{G}_i
\,\frac{\partial}{\partial x_j}(\frac{\partial x_j}{\partial V_i})$.
Utilizing this in (\ref{rho+form}) and integrating by parts
w.r.t. $v$ yields
\begin{eqnarray}\label{zerle+}
   \rho^+(t, x) & = & \int dv\,f^+_0(X(t), V(t))
   -{\rm div}_x\,\int_0^t ds\,\int dv\,\bigg(\frac{\partial x}{\partial V}
   \cdot\tilde{G}\bigg) \nonumber \\ & &
   + \int_0^t ds\,\int dv\,
   \bigg[{\rm div}_x\bigg(\frac{\partial x}{\partial V}\bigg)\cdot\tilde{G}
   + {\rm div}_v\bigg(\frac{\partial v}{\partial V}\bigg)\cdot\tilde{G}\bigg]
   \nonumber \\ & = : & \phi_0^+(t, x)-{\rm div}_x \Gamma^+ (t, x)+R^+(t, x).
\end{eqnarray}
Similarly, we have
\begin{equation}\label{zerle-}
   \rho^-(t, x)=\phi_0^-(t, x)-{\rm div}_x \Gamma^- (t, x)+R^-(t, x),
\end{equation}
with the corresponding functions $\phi_0^-$, $\Gamma^-$, and $R^-$

Next we derive some estimates on $\phi_0^+$, $\Gamma^+$, and $R^+$.

\begin{lemma}\label{estil-1} For $t\in [0, T]$ we have
\[ {\|\phi_0^+(t)\|}_{\frac{3+m}{3};\,x}\le C_T\quad (m>0)\quad\mbox{and}\quad
   {\|\phi_0^+(t)\|}_{3(\frac{3+m}{6+m});\,x}\le C_T\quad (m\ge 3). \]
\end{lemma}
{\bf Proof\,:} We can apply corollary \ref{bd-trafo} below with $s=t$ and $\tau=0$
to obtain the first bound. Concerning the second, note that $m\ge 3$ implies
$3(\frac{3+m}{6+m})\le\frac{3+m}{3}$. Whence it suffices to bound
the support of $x\mapsto\phi_0^+(t, x)=\int dv\,f^+_0(X(t; x, v), V(t; x, v))$.
To do so, recall from (\ref{data-f}) that $f^+_0(\bar{x}, \bar{v})=0$ for
$|\bar{x}|\ge r_0$ or $|\bar{v}|\ge r_0$. {}From the proof of
corollary \ref{bd-trafo} we know $|V(t)-v|\le C_1$, $C_1$ depending only on $T$.
Thus $\frac{\partial}{\partial s}|X(s)-x|\le |V(s)+\eps D^{[2]}(t-s)|
\le C(1+|v|)$ by (\ref{char-syst}) and theorem \ref{dec-summa}(c),
whence $|X(t)-x|\le C_2(1+|v|)$. Then, if $|x|\ge C_2(1+C_1+r_0)+r_0=:r_1$
and $|V(t)|\le r_0$, we have $|v|\le |V(t)-v|+|V(t)|\le C_1+r_0$
and therefore $|X(t)|\ge |x|-|X(t)-x|\ge r_0$. This yields
$f^+_0(X(t), V(t))=0$, and thus $\phi^+_0(t, x)=0$ for $|x|\ge r_1$
and $t\in [0, T]$. {\hfill$\Box$}\bigskip

Next we turn to bound $\Gamma^+(t, x)=\int_0^t ds\,
\int dv\,(\frac{\partial x}{\partial V}\cdot\tilde{G})$.

\begin{lemma}\label{estil-2} For $t\in [0, T]$ and any $t_0\in ]0, T]$ we have
\[ {\|\Gamma^+(t)\|}_{3+m;\,x}\le
   C_T\,t_0^{\,\frac{m-3}{6-m}}\,
   \Big(1+{M_m(t)}^{\frac{9}{(6-m)(3+m)}}\Big)
   +C_T\,(1+|\ln t_0|)\,\Big(1+{M_m(t)}^{\frac{1}{3+m}}\Big), \]
with $3<m<\frac{51}{11}$. Here $C_T$ does not depend on $t_0$. 
Moreover, 
\begin{equation}\label{mifeg} 
   {\|\Gamma^+(t)\|}_{3+m;\,x}\le C_T\,t^{\,\frac{m-3}{6-m}}\,
   \Big(1+{M_m(t)}^{\frac{9}{(6-m)(3+m)}}\Big),\quad t\in [0, T]. 
\end{equation} 
\end{lemma}
{\bf Proof\,:} We first note that
\begin{equation}\label{grap}
   |\Gamma^+(t, x)|\le C_T\int_0^t ds\,s\int dv\,
   \Big|[E_1 f^+](t-s, X(s; x, v), V(s; x, v))\Big|,
\end{equation}
in view of the second estimate in (\ref{detabl-esti}) below.
Next we observe that due to lemma \ref{bd-1} and the first estimate
in (\ref{detabl-esti}) we have, with $\frac{1}{r}+\frac{1}{r'}=1$,
\begin{eqnarray}\label{djrei}
   \lefteqn{ \int dv\,\Big|[E_1 f^+](t-s, X(s), V(s))\Big| }
   \nonumber \\ & \le &
   \sup_{\tau\in [0, T]}{\|f^+(\tau)\|}_{\infty;\,xv}^{\frac{r'-1}{r'}}
   \,{\bigg(\int dv\,{|E_1(t-s, X(s))|}^r\bigg)}^{\frac{1}{r}}
   \,{\bigg(\int dv\,f^+(t-s, X(s), V(s))\bigg)}^{\frac{1}{r'}}
   \nonumber \\ & \le & C_T\,s^{-\frac{3}{r}}\,{\bigg(\int dX
   \,{|E_1(t-s, X)|}^r\bigg)}^{\frac{1}{r}}
   \,{\bigg(\int dv\,f^+(t-s, X(s), V(s))\bigg)}^{\frac{1}{r'}}
   \nonumber \\ & \le &
   C_T\,s^{-\frac{3}{r}}\,\sup_{\tau\in [0, T]}{\|E_1(\tau)\|}_{r;\,x}
   \,{\bigg(\int dv\,f^+(t-s, X(s), V(s))\bigg)}^{\frac{1}{r'}}.
\end{eqnarray}
In (\ref{grap}) we then split $\int_0^t ds=\int_0^{t_0} ds+\int_{t_0}^t ds$
with $t_0\in ]0, t]$.

To bound the first term, we choose $r'=6-m<3$, whence
$r=\frac{6-m}{5-m}>\frac{3}{2}$. Then with $p=\frac{6m-9}{6-m}>1$
we find $\frac{3+m}{r'}=\frac{3+p}{3}$. Moreover, $1\le r<\frac{15}{4}$
by the choice of $m$. Hence by lemma \ref{e1f-lem} and corollary \ref{bd-trafo}
\begin{eqnarray*}
   \lefteqn{ {\bigg\|\int_0^{t_0} ds\,s\int dv\,
   \Big|[E_1 f^+](t-s, X(s; \cdot, v), V(s; \cdot, v))
   \Big|\bigg\|}_{3+m;\,x} }
   \\ & \le & C_T\,t_0^{2-\frac{3}{r}}\,
   \bigg(\sup_{\tau\in [0, T]}{\|E_1(\tau)\|}_{r;\,x}\bigg)
   \\ & & \hspace{5em}\times
   \,\sup_{\tau\in [0, t]}\Bigg[\int dx\,{\bigg(\int dv
   \,f^+(t-\tau, X(\tau), V(\tau))\bigg)}^{\frac{3+m}{r'}}\Bigg]^{\frac{1}{3+m}}
   \\ & \le & C_T\,t_0^{2-\frac{3}{r}}\,\bigg(1+{M_p(t)}^{\frac{1}{3+m}}\bigg);
\end{eqnarray*}
note that Jensen's inequality has been used for the first estimate.
Utilizing the relation $3(\frac{3+2p}{6+p})=m$ and (\ref{mome-mome}),
we can bound $M_p(t)$ by means of $M_m(t)$, as
\[ M_p(t)\le C_T\Big(1+{M_m(t)}^{\frac{6+p}{3}}\Big)
  =C_T\Big(1+{M_m(t)}^{\frac{9}{6-m}}\Big). \]
Thus we have shown that
\begin{eqnarray}\label{befri1}
   \lefteqn{{\bigg\|\int_0^{t_0} ds\,s\int dv\,
   \Big|[E_1 f^+](t-s, X(s; \cdot, v), V(s; \cdot, v))\Big|\bigg\|}_{3+m;\,x}}
   \nonumber \\ & & \hspace{3em}\le C_T\,t_0^{2-\frac{3}{r}}\,
   \Big(1+{M_m(t)}^{\frac{9}{(6-m)(3+m)}}\Big).
\end{eqnarray}

As far as the second part of the integral is concerned, we now make use
of (\ref{djrei}) with $r'=3$ and $r=\frac{3}{2}$. Then
\begin{eqnarray}\label{befri2}
   \lefteqn{ {\bigg\|\int_{t_0}^t ds\,s\int dv\,
   \Big|[E_1 f^+](t-s, X(s; \cdot, v), V(s; \cdot, v))\Big|\bigg\|}_{3+m;\,x} }
   \nonumber \\ & \le &
   C_T\,(1+|\ln t_0|)\,\Big(\sup_{\tau\in [0, T]}{\|E_1(\tau)\|}
   _{\frac{3}{2};\,x}\Big)
\\ & & \hspace{8em}\times
   \,\sup_{\tau\in [0, t]}\Bigg[\int dx\,{\bigg(\int dv
   \,f^+(t-\tau, X(\tau), V(\tau))\bigg)}^{\frac{3+m}{3}}\Bigg]^{\frac{1}{3+m}}
   \nonumber \\ & \le &
   C_T\,(1+|\ln t_0|)\,\Big(1+{M_m(t)}^{\frac{1}{3+m}}\Big),
\end{eqnarray}
again by lemma \ref{e1f-lem} and corollary \ref{bd-trafo}.
Summarizing (\ref{befri1}) and (\ref{befri2}), we see that the first asserted 
estimate holds. To verify (\ref{mifeg}) it sufficient to follow the argument 
just elaborated and to note that for $t_0=t$ the contribution of the 
$\int_{t_0}^t ds(\ldots)$-part of (\ref{grap}) drops out, 
whence we simply use (\ref{befri1}) for $t_0=t$. 
{\hfill$\Box$}\bigskip

Finally we need to consider
\[ R^+(t, x)=\int_0^t ds\,\int dv\,
   \bigg[{\rm div}_x\bigg(\frac{\partial x}{\partial V}\bigg)\cdot\tilde{G}
   + {\rm div}_v\bigg(\frac{\partial v}{\partial V}\bigg)\cdot\tilde{G}\bigg] \]
in (\ref{zerle+}).

\begin{lemma}\label{estil-3} For $t\in [0, T]$ we have
\[ {\|R^+(t)\|}_{3(\frac{3+m}{6+m});\,x}\le C_T, \]
with $m\in [0, \frac{147}{16}]$.
\end{lemma}
{\bf Proof\,:} Using (\ref{2nd-esti}) below, and (\ref{djrei})
with $r'=\frac{13}{9}$ and $r=\frac{13}{4}$, we estimate
\begin{eqnarray*}
   |R^+(t, x)| & \le & C_T\int_0^t ds\,\int dv\,
   \Big|[E_1 f^+](t-s, X(s; x, v), V(s; x, v))\Big| \\ & \le & C_T
   \,\bigg(\sup_{\tau\in [0, T]}{\|E_1(\tau)\|}_{\frac{13}{4};\,x}\bigg)
   \,\int_0^t ds\,s^{-\frac{12}{13}}\,{\bigg(
   \int dv\,f^+(t-s, X(s), V(s))\bigg)}^{\frac{9}{13}}
   \\ & \le & C_T\,\int_0^t ds\,s^{-\frac{12}{13}}
   \,{\bigg(\int dv\,f^+(t-s, X(s), V(s))\bigg)}^{\frac{9}{13}},
\end{eqnarray*}
the latter according to lemma \ref{e1f-lem}. Hence due to corollary
\ref{bd-trafo}, with $p$ determined through
$\frac{3+p}{3}=\frac{27}{13}(\frac{3+m}{6+m})$,
\begin{eqnarray*}
   {\|R^+(t)\|}_{3(\frac{3+m}{6+m});\,x}
   & \le & C_T\sup_{\tau\in [0, t]}\Bigg[\int dx\,{\bigg(\int dv
   \,f^+(t-\tau, X(\tau), V(\tau))\bigg)}
   ^{\frac{27}{13}(\frac{3+m}{6+m})}\Bigg]^{\frac{6+m}{3(3+m)}}
   \\ & \le & C_T\bigg(1+{M_p(t)}^{\frac{6+m}{3(3+m)}}\bigg)
   =C_T\bigg(1+{M_{3(\alpha-1)}(t)}^{\frac{6+m}{3(3+m)}}\bigg),
\end{eqnarray*}
where $\alpha=\frac{27}{13}(\frac{3+m}{6+m})$. Since $0\le 3(\alpha-1)
=\frac{3}{13}(\frac{3+14m}{6+m})\le 2$ by choice of $m$,
the claim follows from lemma \ref{bd-1}.
{\hfill$\Box$}\bigskip

The foregoing estimates, and analogous ones
for $\phi_0^-$, $\Gamma^-$, and $R^-$, can be put together
to yield the following result.

\begin{lemma}\label{holda} For $t\in [0, T]$ and $m\in ]3, \frac{51}{11}[$
we have
\begin{eqnarray*} 
   {\|\nabla U(t)\|}_{3+m;\,x} & \le & C_T\,t_0^{\,\frac{m-3}{6-m}}\,
   \Big(1+{M_m(t)}^{\frac{9}{(6-m)(3+m)}}\Big)
\\ & & 
   +\,C_T(1+|\ln t_0|)\,\Big(1+{M_m(t)}^{\frac{1}{3+m}}\Big),
\end{eqnarray*} 
with $t_0\in ]0, t]$ being arbitrary. Here $C_T$ does not depend on $t_0$. 
Moreover, 
\begin{equation}\label{rockef} 
   {\|\nabla U(t)\|}_{3+m;\,x}\le C_T\,t^{\,\frac{m-3}{6-m}}\,
   \Big(1+{M_m(t)}^{\frac{9}{(6-m)(3+m)}}\Big),\quad t\in [0, T]. 
\end{equation} 
\end{lemma}
{\bf Proof\,:} Due to (\ref{E-def}), (\ref{zerle+}),
and (\ref{zerle-}) we may write
\begin{eqnarray*} 
   \nabla U(t, x) & = & -\bigg(\nabla\frac{1}{|x|}\bigg)\ast
   \Big([\phi_0^+(t)-\phi_0^-(t)]+[R^+(t)-R^-(t)]
\\ & & 
   +\,{\rm div}_x[\Gamma^-(t)-\Gamma^+(t)]\Big)(x).
\end{eqnarray*} 
Therefore lemma \ref{falt-lem} implies for $m\in [0, \infty[$ that
\begin{eqnarray*}
   {\|\nabla U(t)\|}_{3+m;\,x} & \le & C\Big(
   {\|\phi_0^+(t)\|}_{3(\frac{3+m}{6+m});\,x}
   +{\|\phi_0^-(t)\|}_{3(\frac{3+m}{6+m});\,x}
   \\ & & \hspace{2em} +\,{\|R^+(t)\|}_{3(\frac{3+m}{6+m});\,x}
   +{\|R^-(t)\|}_{3(\frac{3+m}{6+m});\,x}\Big)
   \\ & & +\,C\Big({\|\Gamma^+(t)\|}_{3+m;\,x}
   +{\|\Gamma^-(t)\|}_{3+m;\,x}\Big).
\end{eqnarray*}
Due to lemmas \ref{estil-1}, \ref{estil-2}, and \ref{estil-3}
we thus obtain the first desired estimate. Concerning (\ref{rockef}), 
we rather apply (\ref{mifeg}) than the first estimate from lemma \ref{estil-2}.  
{\hfill$\Box$}\bigskip

This in particular can be used to derive a short time bound on $M_m(t)$.

\begin{cor}\label{loc-cor} For $m\in ]3, \frac{51}{11}[$ there exist $t_1\in ]0, T]$
and $C_1>0$ (both depending on $T$) such that
\[ M_m(t)\le C_1,\quad t\in [0, t_1]. \]
\end{cor}
{\bf Proof\,:} Combining (\ref{Mp-diffint}) with (\ref{rockef}) and observing 
$t^{\frac{m-3}{6-m}}\le C_T$ yields
\begin{eqnarray*} 
   \frac{d}{dt}M_m(t) & \le & C\sup_{s\in [0, t]}\,{\|\nabla U(s)\|}_{3+m;\,x}
   \,{M_m(t)}^{\frac{2+m}{3+m}} \\ & \le & 
   C_T\Big(1+{M_m(t)}^{\frac{9}{(6-m)(3+m)}}\Big)
   \,{M_m(t)}^{\frac{2+m}{3+m}}. 
\end{eqnarray*} 
Integration of this differential inequality gives a local bound on
$M_m(t)$, that, however, fails to extend to all of $[0, T]$ due to
$\frac{9}{(6-m)(3+m)}+\frac{2+m}{3+m}=\frac{7-m}{6-m}>1$. {\hfill$\Box$}\bigskip

\begin{cor}\label{chek} For $t\in [0, T]$ and $m\in ]3, \frac{51}{11}[$
we have
\begin{equation}\label{treas}
   {\|\nabla U(t)\|}_{3+m;\,x}\le
   C_T(1+|\ln M_m(t)|)\,\Big(1+{M_m(t)}^{\frac{1}{3+m}}\Big).
\end{equation}
\end{cor}
{\bf Proof\,:} Note that we may assume $M_m(t)\ge 1$ for $t\in [0, T]$,
since otherwise $M_m(t)$ simply can be replaced by $M_m(t)+1$. Set
\[ t_0=t_1 M_m(t)^{\frac{1}{3-m}}\le t_1 \]
in lemma \ref{holda}, with $t_1$ from corollary \ref{loc-cor}.
If $t\in [t_1, T]$, then $t_0\le t$, and therefore
$\frac{1}{3-m}(\frac{m-3}{6-m})+\frac{9}{(6-m)(3+m)}
=\frac{1}{3+m}$ shows that (\ref{treas}) holds for $t\in [t_1, T]$.
On the other hand, if $t\in [0, t_1]$, then $M_m(t)\le C_1$ and
(\ref{rockef}) imply ${\|\nabla U(t)\|}_{3+m;\,x}\le C_2$ for some
$C_2>0$. Hence (\ref{treas}) holds as well in this case
if we choose $C_T\ge C_2$. {\hfill$\Box$}\bigskip

\begin{theorem}\label{Mm-bd} For $t\in [0, T]$
and $m\in ]3, \frac{51}{11}[$ we have
\[ M_m(t)\le C_T. \]
\end{theorem}
{\bf Proof\,:} By (\ref{Mp-diffint}) and due to corollary \ref{chek}
we see that
\begin{eqnarray*}
   \frac{d}{dt}M_m(t) & \le & C\sup_{s\in [0, t]}\,{\|\nabla U(s)\|}_{3+m;\,x}
   \,{M_m(t)}^{\frac{2+m}{3+m}}\le C_T(1+|\ln M_m(t)|)\,\Big(1+M_m(t)\Big)
   \\ & \le & C_T(1+|\ln M_m(t)|\,M_m(t)\Big).
\end{eqnarray*}
Integration of this differential inequality yields the claimed estimate.
{\hfill$\Box$}\bigskip

\begin{cor}\label{rho-betbd}
For $t\in [0, T]$ we have
\[ {\|\rho^{\pm}(t)\|}_{p;\,x}\le C_T,\quad p\in ]2, \frac{28}{11}[. \]
\end{cor}
{\bf Proof\,:} This is a consequence of (\ref{terrac}) and theorem \ref{Mm-bd},
since $m=3(p-1)\in ]3, \frac{51}{11}[$ corresponds
to $p=\frac{3+m}{3}\in ]2, \frac{28}{11}[$. {\hfill$\Box$}\bigskip

\subsection{Global existence of solutions}

We start with some preliminary (well-known) observations.
Recall the definition of $P^{\pm}(t)$ from (\ref{P-def}),
and also that $[0, T_\ast[$ is the maximal interval of existence,
cf.~theorem \ref{loc-ex}.

\begin{lemma}\label{P-esti} We have
\[ P^{\pm}(t)\le P^{\pm}(0)+\int_0^t {\|\nabla U(s)\|}_{\infty;\,x}
   \,ds,\quad t\in [0, T_\ast[. \]
\end{lemma}
{\bf Proof\,:} Assume e.g.~$(x, v)\in {\rm supp}f^+(s)$
for some $x\in\R^3$ and $s\in [0, t]$. {}From the proof of lemma \ref{bd-1}
we have $f^+(s, x, v)=f^+_0({\cal X}(0; s, x, v), {\cal V}(0; s, x, v))$,
with $({\cal X}, {\cal V})$ the characteristics from (\ref{doz}).
This means that $(x, v)=({\cal X}(s; 0, x_0, v_0),
{\cal V}(s; 0, x_0, v_0))$ for $(x_0, v_0)\in {\rm supp}f^+_0$.
Hence
\[ |v|\le |v_0|+\bigg|\int_0^s\dot{{\cal V}}(\tau; 0, x_0, v_0)\,d\tau\bigg|
   \le P^+(0)+\int_0^t {\|\nabla U(\tau)\|}_{\infty;\,x}\,d\tau \]
by the characteristic equation for ${\cal V}$.
{\hfill$\Box$}\bigskip

Next we need to derive a bound on ${\|\nabla U(t)\|}_{\infty;\,x}$.

\begin{lemma}\label{nabU-esti} For $\alpha>\frac{3}{2}$ we have
\[ {\|\nabla U(t)\|}_{\infty;\,x}\le C
   {\bigg(\sum_{j=\pm}{\|\rho^j(t)\|}_{\infty;\,x}\bigg)}^{1-\frac{\,\alpha'}{3}}
   \,{\bigg(\sum_{j=\pm}{\|\rho^j(t)\|}_{\alpha';\,x}\bigg)}^{\frac{\,\alpha'}{3}},
   \quad t\in [0, T_\ast[, \]
where $\frac{1}{\alpha}+\frac{1}{\alpha'}=1$. The constant $C$
depends only on $\alpha$.
\end{lemma}
{\bf Proof\,:} With $R>0$ we estimate from (\ref{E-def})
\begin{eqnarray*}
   |\nabla U(t, x)| & \le & \int_{|y-x|\le R}\,\frac{dy}{|x-y|^2}
   \Big(\rho^+(t, y)+\rho^-(t, y))\Big)
   \\ & & +\,\int_{|y-x|\ge R}\,\frac{dy}{|x-y|^2}
   \Big(\rho^+(t, y)+\rho^-(t, y))\Big)
   \\ & \le &
   C\bigg(\sum_{j=\pm}{\|\rho^j(t)\|}_{\infty;\,x}\bigg)R
   +{\bigg(\int_{|y-x|\ge R}\,\frac{dy}{|x-y|^{2\alpha}}
   \bigg)}^{\frac{1}{\alpha}}\,\bigg(\sum_{j=\pm}\,{\|\rho^j(t)\|}_{\alpha';\,x}\bigg)
   \\ & \le & C\bigg(\sum_{j=\pm}{\|\rho^j(t)\|}_{\infty;\,x}\bigg)R
   +C\bigg(\sum_{j=\pm}{\|\rho^j(t)\|}_{\alpha';\,x}\bigg)R^{\,\frac{3}{\alpha}-2}.
\end{eqnarray*}
Choosing the optimal $R$ yields the claim.
{\hfill$\Box$}\bigskip

\begin{lemma}\label{rho-P} We have
\[ {\|\rho^{\pm}(t)\|}_{\infty;\,x}\le C P^{\pm}(t)^3,\quad t\in [0, T_\ast[, \]
with $C$ depending only on the data.
\end{lemma}
{\bf Proof\,:} By definition of $P^+(t)$ and bounding
${\|f^{\pm}(t)\|}_{\infty;\,xv}$ as in lemma \ref{bd-1}, it follows that
\[ \rho^+(t, x)=\int_{|v|\le P^+(t)}f^+(t, x, v)\,dv
   \le C{\|f^{\pm}(t)\|}_{\infty;\,xv}P^+(t)^3\le CP^+(t)^3, \]
as was to be shown.
{\hfill$\Box$}\bigskip

Using the criterion from theorem \ref{loc-ex}
and by means of corollary \ref{rho-betbd}
we are finally going to complete the proof of theorem \ref{glob-ex}. \medskip

\noindent
{\bf Proof of theorem \ref{glob-ex}\,:} Assume $T_\ast<\infty$ in theorem \ref{loc-ex}.
All the estimates on the moments remain valid if $[0, T]$ is replaced
by $[0, T_\ast[$, since in the constants only terms of the form
$CT_\ast$, $T_\ast^\alpha$, and $e^{CT_\ast}$ do enter.
In particular,
\[ {\|\rho^{\pm}(t)\|}_{\frac{27}{11};\,x}\le C,\quad t\in [0, T_\ast[, \]
by corollary \ref{rho-betbd}, where here and below the various constants $C$
depend on $T_\ast$. Choosing $\alpha=\frac{27}{16}>\frac{3}{2}$,
which corresponds to $\alpha'=\frac{27}{11}$, we deduce
from lemmas \ref{P-esti}, \ref{nabU-esti}, and \ref{rho-P} that
\begin{eqnarray*}
   P^+(t) & \le & P^+(0)+C\int_0^t ds
   \,{\bigg(\sum_{j=\pm}{\|\rho^j(s)\|}_{\infty;\,x}\bigg)}^{\frac{2}{11}}
   \,{\bigg(\sum_{j=\pm}{\|\rho^j(s)\|}_{\frac{27}{11};\,x}\bigg)}^{\frac{9}{11}}
   \\ & \le & P^+(0)+C\int_0^t ds\,{\Big(P^+(s)^3+P^-(s)^3\Big)}^{\frac{2}{11}}.
\end{eqnarray*}
This implies
\[ P(t)\le P(0)+C\int_0^t P(s)^{\frac{6}{11}}\,ds,\quad t\in [0, T_\ast[, \]
and hence the boundedness of $P$ on $[0, T_\ast[$.
{\hfill$\Box$}\bigskip

We remark that we did not try to reduce the exponent $\frac{6}{11}$ so as
to obtain the optimal power of $P$.

\subsection{Some technical lemmas}

\begin{lemma}\label{e1f-lem} Define $E_1$ and $F$ as in (\ref{E1-def}).
Then we have ${\|E_1(t)\|}_{p;\,x}\le C$ for $t\in [0, T]$
and $p\in [1, \frac{15}{4}[$, as well as
\begin{equation}\label{F-esti}
   {\|F(t)\|}_{\infty;\,x}+{\|\nabla F(t)\|}_{\infty;\,x}
   +{\|D^2 F(t)\|}_{\infty;\,x}\le CR^{-2},\quad t\in [0, T].
\end{equation}
\end{lemma}
{\bf Proof\,:} {}From (\ref{E1-def}) and Young's inequality \cite[p.~29]{reedsimon2}
with $\frac{1}{q}+\frac{1}{r}=1+\frac{1}{p}$ we obtain
\[ {\|E_1(t)\|}_{p;\,x}
   ={\bigg\|(\chi\,\nabla\frac{1}{|x|})\ast (\rho^+(t)-\rho^-(t))\bigg\|}_{p;\,x}
   \le {\bigg\|\chi\,\nabla\frac{1}{|x|}\bigg\|}_{q;\,x}
   {\|\rho^+(t)-\rho^-(t)\|}_{r;\,x}. \]
We have $\chi(\cdot)\,\frac{x}{|x|^3}\in L^q(\R^3)$ for $q\in [1, \frac{3}{2}[$
and ${\|\rho^{\pm}(t)\|}_{r;\,x}\le C$ for $r\in [1, \frac{5}{3}]$
due to theorem \ref{dec-summa}(a). Combining those values for $q$ and $r$,
we see that we need to have $p\in [1, \frac{15}{4}[$. The bounds in (\ref{F-esti})
are obtained by observing that
\[ F(t, x)=\int_{|x-y|\ge R}\Big(1-\chi(x-y)\Big)
   \frac{(x-y)}{|x-y|^3}(\rho^+(t, y)-\rho^-(t, y))\,dy \]
where $\chi\in C_0^\infty(\R^3)$, and moreover $\int\rho^{\pm}(t, y)\,dy
=\int\int f^{\pm}(t, y, v)\,dydv\le C$. {\hfill$\Box$}\bigskip

\begin{lemma}\label{trafo} For fixed $t\in ]0, T]$ and $s\in [0, t]$
consider the map
\[ Z(s):\quad\R^6\ni (x, v)\mapsto (X(s; x, v), V(s; x, v))
   =(X(s), V(s))\in\R^6, \]
where $(X(s), V(s))$ is the solution of the characteristic system
(\ref{char-syst}), i.e.
\[ \left(\begin{array}{c} \dot{X}(s) \\ \dot{V}(s)\end{array}\right)
   =\left(\begin{array}{c} -V(s)-\eps D^{[2]}(t-s)
   \\ -F(t-s, X(s))\end{array}\right), \quad
   \left(\begin{array}{c} X(0) \\ V(0)\end{array}\right)
   =\left(\begin{array}{c} x \\ v\end{array}\right). \]
Then $Z(s)$ is a volume-preserving diffeomorphism, and
\begin{equation}\label{detabl-esti}
   \bigg|\det\bigg(\frac{\partial X}{\partial v}(s)\bigg)^{-1}\bigg|
   \le C s^{-3},\quad \bigg|\frac{\partial x}{\partial V}(s)\bigg|\le C s,
\end{equation}
as well as
\begin{equation}\label{2nd-esti}
   \bigg|\frac{\partial}{\partial x_i}
   \bigg(\frac{\partial x_j}{\partial V_k}(s)
   \bigg)\bigg|+\bigg|\frac{\partial}{\partial v_i}
   \bigg(\frac{\partial v_j}{\partial V_k}(s)\bigg)\bigg|\le
   C,\quad 1\le i, j, k\le 3,
\end{equation}
if $R>0$ is chosen sufficiently large (depending on $T$).
Here
\[ Z(s)^{-1}: (X, V)\mapsto (x, v)=(x(s; X, V), v(s; X, V)) \]
is the inverse of $Z(s)$, and the constants $C$ do depend only on $T$,
but not on $(t, s, x, v)$.
\end{lemma}
{\bf Proof\,:} As the right-hand side of the characteristic
system has divergence ${\rm div}={\rm div}_{(X, V)}=0$, the
first claim follows; cf.~also remark \ref{bigsav}. Moreover, we have
\begin{eqnarray}
   & & \bigg|\frac{\partial X}{\partial v}(s)+s\,{\rm Id}\bigg|\le
   C R^{-2} s^3,\quad
   \bigg|\frac{\partial V}{\partial v}(s)-{\rm Id}\bigg|\le
   C R^{-2} s^2, \label{johta1} \\
   & & \bigg|\frac{\partial X}{\partial x}(s)-{\rm Id}\bigg|\le
   C R^{-2} s^2,\quad
   \bigg|\frac{\partial V}{\partial x}(s)\bigg|\le C R^{-2} s,
   \label{johta2}
\end{eqnarray}
where ${\rm Id}$ denotes the unit matrix in $\R^3$. E.g.~to validate
the estimate on the $v$-derivatives one can introduce, following \cite{bar-deg},
the function $\phi(s)=\frac{\partial X}{\partial v}(s)+s\,{\rm Id}$ and
calculate that $\phi(0)=0$, $\dot{\phi}(s)=-\frac{\partial V}{\partial v}(s)+{\rm Id}$,
$\dot{\phi}(0)=0$, as well as 
\[ \ddot{\phi}(s)=\frac{\partial F}{\partial x}(t-s,
X(s))
   [\phi(s)-s\,{\rm Id}]. \] 
Here the important observation is that
\begin{eqnarray*} 
   \dot{\phi}(s) & = & \frac{\partial}{\partial s}
   \bigg(\frac{\partial X}{\partial v}(s)\bigg)
   +{\rm Id}=\frac{\partial}{\partial v}\dot{X}(s)+{\rm Id}
   =\frac{\partial}{\partial v}\bigg(-V(s)-\eps D^{[2]}(t-s)\bigg)
   +{\rm Id} \\ & = & -\,\frac{\partial V}{\partial v}(s)+{\rm Id}, 
\end{eqnarray*}
since the term with $\eps D^{[2]}(t-s)$ simply drops through the $v$-derivative,
and thus the same general argument can be used as in the case without $D^{[2]}$.
Then we may write $\phi(s)=\int_0^s (s-\tau)\ddot{\phi}(\tau)\,d\tau$
and utilize (\ref{F-esti}) to derive (\ref{johta1}) by means of a Gronwall argument,
whereas (\ref{johta2}) is obtained in the same way using the
function $\phi(s)=\frac{\partial X}{\partial x}(s)-{\rm Id}$ instead.

Then
\[ \bigg|\det\bigg(\frac{\partial X}{\partial v}(s)\bigg)\bigg|
   =s^3\,\bigg|\det\bigg(-s^{-1}\bigg[\frac{\partial X}{\partial v}(s)
   +s\,{\rm Id}\bigg]+{\rm Id}\bigg)\bigg| \]
together with (\ref{johta1}) and the continuity of the map $A\mapsto |\det(A)|$
at $A={\rm Id}$ yields that for $s\in [0, T]$ and $R>0$ large enough,
\[ \bigg|\det\bigg(\frac{\partial X}{\partial v}(s)\bigg)\bigg|
   \ge\frac{1}{2}s^3, \]
and this proves the first estimate in (\ref{detabl-esti}).
As what concerns the bound on $|\frac{\partial x}{\partial V}(s)|$,
employing the chain rule it follows that
\begin{eqnarray*} 
   \frac{\partial x}{\partial V}(s) & = & 
   -\,\bigg(\frac{\partial x}{\partial X}(s)\bigg)
   \bigg(\frac{\partial X}{\partial v}(s)\bigg)
   {\bigg(\frac{\partial V}{\partial v}(s)\bigg)}^{-1}, 
   \\ \frac{\partial x}{\partial X}(s) & = & \bigg[{\rm Id}
   -\bigg(\frac{\partial x}{\partial V}(s)\bigg)
   \bigg(\frac{\partial V}{\partial x}(s)\bigg)\bigg]
   {\bigg(\frac{\partial X}{\partial x}(s)\bigg)}^{-1}.
\end{eqnarray*} 
Choosing $R>0$ sufficiently large, we find from (\ref{johta1})
and (\ref{johta2}) that
\[ \bigg|{\bigg(\frac{\partial V}{\partial v}(s)\bigg)}^{-1}\bigg|
   +\bigg|{\bigg(\frac{\partial X}{\partial x}(s)\bigg)}^{-1}\bigg|\le C, \]
whence by (\ref{johta1}) and (\ref{johta2}),
\[ \bigg|\frac{\partial x}{\partial V}(s)\bigg|
   \le Cs\,\bigg|\frac{\partial x}{\partial X}(s)\bigg|,\quad
   \bigg|\frac{\partial x}{\partial X}(s)\bigg|
   \le C\bigg(1+R^{-2}s\,\bigg|\frac{\partial x}{\partial V}(s)\bigg|\bigg), \]
and this gives $|\frac{\partial x}{\partial V}(s)|\le Cs$,
for $R>0$ large enough.

Finally, the estimates on the second derivatives in (\ref{2nd-esti})
are more tedious, but verified in a similar way.
{\hfill$\Box$}\bigskip

\begin{cor}\label{bd-trafo} If $(X(s), V(s))$ is a characteristic curve,
cf.~lemma \ref{trafo}, and $s, \tau\in [0, T]$, then for $p\in ]0, \infty[$
\[ {\bigg\|\int f^{\pm}(\tau, X(s; \cdot, v), V(s; \cdot, v))\,dv
   \bigg\|}_{\frac{3+p}{3};\,x}\le C_T\,{\|f^{\pm}(\tau)\|}_{\infty;\,xv}^{\frac{p}{3+p}}
   \,\Big({M_0(\tau)}^{\frac{3}{3+p}}+{M_p(\tau)}^{\frac{3}{3+p}}\Big), \]
with $M_p(\cdot)$ from (\ref{mom-def}).
\end{cor}
{\bf Proof\,:} Utilizing lemma \ref{R-arg} with $f(x, v)
=f^{\pm}(\tau, X(s; x, v), V(s; x, v))$, it follows that
the left-hand side is bounded by
\[ C\,{\|f^{\pm}(\tau)\|}_{\infty;\,xv}^{\frac{p}{3+p}}
   \,{\bigg(\int\int\,|v|^p\,f^{\pm}(\tau, X(s; x, v), V(s; x, v))
   \,dxdv\bigg)}^{\frac{3}{3+p}}. \]
{}From the characteristic equation for $V(s)$ we obtain that $\frac{\partial}{\partial s}
|V(s)-v|\le {\|F\|}_{\infty;\,xt}\le C$, whence $|V(s)-v|\le CT=C_T$, thus
${|v|}^p\le C(1+{|V(s)|}^p)$. Using this estimate and the fact
that $Z(s)$ is a volume-preserving diffeomorphism yields the claim.
{\hfill$\Box$}


\bigskip

\begin{thebibliography}{99}

\bibitem{adams}
 Adams R.A., {\em Sobolev Spaces}, Academic Press,
 New York 1975
\bibitem{andre}
 Andr\'{e}asson H., Global existence of smooth solutions in three
 dimensions for the semiconductor Vlasov-Poisson-Boltzmann equation,
 {\em Nonlinear Anal.~}{\bf 28}, 1193-1211 (1997)
\bibitem{bar-deg}
 Bardos C.\,\& Degond P., Global existence for the Vlasov-Poisson
 equation in $3$ space variables with small initial data,
 {\em Ann.~Inst.~H.~Poincar\'{e} Anal.~Non Lin\'{e}aire}{\bf\,\,2},
 101-118 (1985)
\bibitem{batt}
 Batt J., Global symmetric solutions of the initial value problem in
 stellar dynamics, {\em J.~Differential Equations}{\bf\,\,25}, 342-364 (1977)
\bibitem{bds}
 Blanchet L., Damour T.\,\& Sch\"afer G., Post-Newtonian hydrodynamics
 and post-Newtonian gravitational wave generation for numerical relativity,
 {\em Mon.~Not.~R.~Astron.~Soc.~}{\bf 242}, 289-305 (1990)
\bibitem{boc}
 Bouchut F., Existence and uniqueness of a global smooth solution
 for the Vlasov-Poisson-Fokker-Planck system in three dimensions,
 {\em J.~Funct.~Anal.~}{\bf 111}, 239-258 (1993)
\bibitem{burke}
 Burke W.L., Gravitational radiation damping of slowly moving
 systems calculated using matched asymptotic expansions,
 {\em J.~Math.~Phys.~}{\bf 12}, 401-418 (1971)
\bibitem{flanagan}
 Flanagan \'E.\'E., Astrophysical sources of gravitational radiation
 and prospects for their detection, in Dadhich N.\,\& Narlikar J.~(Eds.):
 {\em Gravitation and Relativity at the Turn of the Millenium},
 Inter-University Centre for Astronomy and Astrophysics, Pune 1998
\bibitem{glassey}
 Glassey R.T., {\em The Cauchy Problem in Kinetic Theory},
 SIAM, Philadelphia 1996
\bibitem{horm}
 H\"ormander L., {\em The Analysis of Linear
 Partial Differential Operators I}, Springer, Berlin-New York 1983
\bibitem{horst}
 Horst E., On the asymptotic growth of the solutions
 of the Vlasov-Poisson system, {\em Math.~Meth.~Appl.~Sci.~}{\bf 16},
 75-85 (1993)
\bibitem{illner-rein}
 Illner R.\,\& Rein G., Time decay of the solutions of the
 Vlasov-Poisson system in the plasma physical case,
 {\em Math.~Meth.~Appl.~Sci.~}{\bf 19}, 1409-1413 (1996)
\bibitem{jackson}
 Jackson J.D., {\em Classical Electrodynamics}, Wiley, New York 1975
\bibitem{KS1}
 Kunze M.\,\& Spohn H., Radiation reaction and center
 manifolds, {\em SIAM J.~Math.~Anal.~}{\bf 32}, 30-53 (2000)
\bibitem{KS2}
 Kunze M.\,\& Spohn H., Adiabatic limit for the Maxwell-Lorentz
 equations, {\em Annales H.~Poincar\'{e}}{\bf\,\,1}, 625-653 (2000)
\bibitem{Nteil}
 Kunze M.\,\& Spohn H., Slow motion of charges interacting
 through the Maxwell field, {\em Comm.~Math.~Phys.~}{\bf 212},
 437-467 (2000)
\bibitem{lio-pert}
 Lions P.L.\,\& Perthame B., Propagation of moments and regularity
 for the $3$-dimensional Vlasov-Poisson system,
 {\em Invent.~Math.~}{\bf 105}, 415-430 (1991)
\bibitem{perthame}
 Perthame B., Time decay, propagation of low moments
 and dispersive effects for kinetic equations,
 {\em Comm.~Partial Differential Equations}{\bf\,\,21}, 659-686 (1996)
\bibitem{pfaff}
 Pfaffelmoser K., Global classical solutions of the Vlasov-Poisson
 system in three dimensions for general initial data,
 {\em J.~Differential Equations}{\bf\,\,95}, 281-303 (1992)
\bibitem{pulvi}
 Pulvirenti M.\,\& Simeoni C., $L^\infty$-estimates for the
 Vlasov-Poisson-Fokker-Planck equation,
 {\em Math.~Meth.~Appl.~Sci.~}{\bf 23}, 923-935 (2000)
\bibitem{reedsimon2}
 Reed M.\,\& Simon B., {\em Methods of Modern Mathematical
 Physics II: Fourier Analysis, Self Adjointness}, Academic Press,
 New York 1975
\bibitem{gerh}
 Rein G., Selfgravitating systems in Newtonian theory -- the
 Vlasov-Poisson system, in {\em Proc.~Minisemester on Math.~Aspects
 of Theories of Gravitation 1996}, Banach Center Publications {\bf 41},
 part I, 179-194 (1997)
\bibitem{rein-rendall}
 Rein G.\,\& Rendall A., Global existence of classical solutions to
 the Vlasov-Poisson system in a three dimensional, cosmological setting,
 {\em  Arch.~Rat.~Mech.~Anal.~}{\bf 126}, 183-201 (1994)
\bibitem{schaeff}
 Schaeffer J., Global existence of smooth solutions to the
 Vlasov-Poisson system in three dimensions,
 {\em Comm.~Partial Differential Equations}{\bf\,\,16}, 1313-1335 (1991)
\bibitem{zhelez}
 Zheleznyakov V.V., {\em Radiation in Astrophysical Plasmas},
 Kluwer, Dordrecht 1996

\end{thebibliography}
\end{document}